\documentclass[twocolumn,tighten]{aastex61}
\usepackage{amsmath,amstext}
\usepackage[T1]{fontenc}
\usepackage{apjfonts} 
\usepackage[figure,figure*]{hypcap}

 \def\mso{\,\mathrm{M}_\odot}

 \def\simle{\mathrel{\hbox{\rlap{\hbox{\lower4pt\hbox{$\sim$}}}\hbox{$<$}}}}
 \def\simgr{\mathrel{\hbox{\rlap{\hbox{\lower4pt\hbox{$\sim$}}}\hbox{$>$}}}}

\shorttitle{Progenitor models for lGRBs and Type\,Ic SLSNe}
\shortauthors{Aguilera-Dena et al.}

\begin{document}

\title{Related progenitor models for long-duration gamma ray bursts and Type\,Ic superluminous supernovae}

\author{David R. Aguilera-Dena}
\affiliation{Argelander-Institut f\"ur Astronomie, Universit\"at Bonn, Auf dem H\"ugel 71, 53121 Bonn, Germany}
\author{Norbert Langer}
\affiliation{Argelander-Institut f\"ur Astronomie, Universit\"at Bonn, Auf dem H\"ugel 71, 53121 Bonn, Germany}
\affiliation{Max-Planck-Institut f\"ur Radioastronomie, Auf dem H\"ugel 69, 53121 Bonn, Germany}
\author{Takashi J. Moriya}
\affiliation{Division of Theoretical Astronomy, National Astronomical Observatory of Japan, National Institutes for Natural Sciences,2-21-1 Osawa, Mitaka, Tokyo 181-8588, Japan}
\author{Abel Schootemeijer}
\affiliation{Argelander-Institut f\"ur Astronomie, Universit\"at Bonn, Auf dem H\"ugel 71, 53121 Bonn, Germany}
\correspondingauthor{David R. Aguilera-Dena}
\email{davidrad@astro.uni-bonn.de}

\begin{abstract}

We model the late evolution and mass loss history of rapidly rotating Wolf-Rayet stars
in the mass range $5\,\rm{M}_{\odot}\dots 100\,\rm{M}_{\odot}$. 
We find that quasi-chemically homogeneously evolving single stars computed with enhanced mixing
retain very little or no helium and are compatible with Type\,Ic supernovae.
The more efficient removal of core angular momentum and the expected smaller compact object mass
in our lower mass models 
lead to core spins in the range suggested for magnetar driven superluminous supernovae.
Our more massive models retain larger specific core angular momenta, expected for
long-duration gamma-ray bursts in the collapsar scenario.
Due to the absence of a significant He envelope,
the rapidly increasing neutrino emission after core helium exhaustion 
leads to an accelerated contraction of the whole star,
inducing a strong spin-up, and centrifugally driven
mass loss at rates of up to $10^{-2}\,\rm{M}_{\odot}~\rm{yr^{-1}}$ in the last years to decades before core collapse. 
Since the angular momentum transport in our lower mass models enhances the envelope spin-up, they show the largest
relative amounts of centrifugally enforced mass loss, i.e., up to 25\% of the expected ejecta mass.
Our most massive models evolve into the pulsational pair-instability regime. 
We would thus expect signatures of interaction with a C/O-rich circumstellar medium
for Type~Ic superluminous supernovae with ejecta masses below $\sim 10\,\rm{M}_{\odot}$ and for the most massive engine-driven explosions with ejecta masses above $\sim 30\,\rm{M}_{\odot}$.  
Signs of such interaction should be observable at early epochs of the supernova explosion, and may be
related to bumps observed in the light curves of superluminous supernovae, or to the massive circumstellar CO-shell proposed
for Type~Ic superluminous supernova Gaia16apd.
\end{abstract}


\section{Introduction}

With the advent of multi-wavelength, all-sky surveys dedicated to observing the onset and aftermath of transient astrophysical events such as supernovae (SNe) and gamma-ray bursts (GRBs), and the fast follow-up observations carried out by space-based and ground instruments, our understanding of the properties of their progenitors and environments has dramatically increased in the last decades (e.g., \citealt{2009PASP..121.1334R,2009ApJ...696..870D}; \citealt{2010SPIE.7733E..0EK}; \citealt{2015AJ....150..150F}; \citealt{2016ApJ...819....5T}).
As observations become more numerous and have higher resolution and time coverage, it becomes possible to refine our theoretical understanding of the environment and physical conditions that led to these events.

The large increase in the number of observed SNe and the improvement in the analysis techniques has also led to identify new classes 
of SNe, well beyond the classical Types Ia, Ib, Ic, and II. In particular, two of these new classes are currently thought to
be caused by rapidly rotating H- and He-poor massive progenitor stars, the broad-lined Type\,Ic SNe (SNe\,Ic-bl, also known as
hypernovae, HNe), and the superluminous Type\,Ic SNe (SLSNe-Ic). A fraction of the SNe\,Ic-bl is observationally associated with
long-duration gamma-ray bursts (lGRBs) (\citealt{2016ApJ...832..108M}), which have been suggested to be triggered by the formation of rapidly rotating accreting black holes in the collapsar scenario \citep{Woosley1993}, or by formation of rapidly spinning super-magnetic neutron stars in the
magnetar scenario \citep{1992Natur.357..472U,2010ApJ...719L.204W}.
While conceivably all lGRBs are associated with the supernova explosion of a rapidly rotating
Wolf-Rayet star, 
not all SNe\,Ic-bl appear to be associated with an lGRB \citep{2006Natur.444.1047F,2006Natur.444.1044G,2006Natur.444.1050D,2006Natur.444.1053G}. Nevertheless, their large explosion energies and large-scale asphericity \citep{2010NewAR..54..191N} demand an explosion mechanism other than that of ordinary core-collapse SNe,
which might be provided by the formation of a central rapidly rotating compact object. 

Whereas detailed progenitor evolution models for SLSNe-Ic are yet to be developed,
\cite{2017ApJ...850...55N}, based on the analysis of a large sample 
of observed SLSNe, suggest that they may be products of 
chemically homogeneous evolution, or were spun up through binary interaction.
It appears that the magnetar scenario can
explain the required large on timescales of weeks
as being tapped from the rotational energy of a millisecond magnetar \citep{Kasen2010,2010ApJ...719L.204W}. 
Energy input on these timescales is provided in ordinary supernovae through the
decay of radioactive nickel. However, the large amounts of nickel required to explain the peak brightness appears to be incompatible with many Type\,Ic SLSN observations \citep[e.g.,][]{2013ApJ...770..128I,2013Natur.502..346N}. 
Whereas the necessary B-field in this case is somewhat smaller than the one needed to produce lGRBs (\citealt{2015MNRAS.454.3311M}; \citealt{2018MNRAS.475.2659M})
also here a nearly critically rotating Wolf-Rayet star is indispensable for producing the magnetar at core collapse.

Several studies have accounted for the effect of rotation in massive star evolution models, highlighting its impact on observable 
stellar properties, including the possible consequences that the high rotation rate may have for the ultimate fate of the star  
\citep{HegerLangerWoosley2000, 2012A&A...537A.146E, 2012A&A...542A..29G,2013A&A...558A.103G, Yoon2006, Brott2011, ChatzopolousWheeler2012, Szecsi2015}. 
In particular, chemically homogeneous evolution (CHE)  due to rotationally induced mixing, which is favoured to occur at low metallicities, 
has been suggested to lead to the formation of lGRBs \citep{2005A&A...443..643Y,2006ApJ...637..914W}, but also various
close binary evolution channels have been suggested to produce rapidly rotating Wolf-Rayet stars at core collapse leading to
lGRBs and SNe\,Ic-bl \citep{2008A&A...484..831D,2010MNRAS.406..840P}.

In this paper, we investigate the pre-collapse evolution of rapidly rotating stars devoid of H, with an
emphasis on the the structural changes which occur after core He exhaustion.
In Sect.\,\ref{sec:method} we describe the method we used to compute our 
stellar evolution sequences, and the assumptions that go into our calculations. In Sect.\,\ref{sec:results} we present our results, 
while in Sect.\,\ref{sec:consequences} we present possibly observable consequences of our findings in SNe. 
We end the paper with a brief discussion in Sect.\,\ref{sec:discussion} before presenting our conclusions in Sect.\,\ref{sec:conclusions}. 

\section{Method} \label{sec:method}

We performed evolutionary calculations of rotating massive stars using the Modules for Experiments in Stellar Astrophysics (MESA) code --in its version r10000-- which includes the effects of stellar rotation \citep{MESAI,MESAII,MESAIII,2017arXiv171008424P}. Our initial models have uniform composition, and a metallicity of Z = 0.00034, or $\sim$ 1/50 Z$_{\odot}$, scaled from solar metallicity \citep{1996ASPC...99..117G}, and an initial helium mass fraction of Y=0.2477  \citep{2007ApJ...666..636P}. While this metallicity does not
represent the average of the observed SLSNe\,Ic population, it was chosen to avoid drastic spin-down
and maximize the effects obtained from rotationally induced chemically homogeneous evolution.
We provide a discussion of the consequences of our choice in Sect.\,\ref{sec:discussion}.

The initial masses of the evolutionary sequences we discuss are between 5 and 100 $\mso$, and they an have initial rotational velocity of 600 km/s, which corresponds to a fraction of their critical rotation rate in the range of 0.68 to 0.89. This velocity corresponds
to that of the fastest rotating known O\,star \citep{2013A&A...560A..29R}, and is chosen such that 
rotationally induced chemically homogeneous evolution is guaranteed \citep{Yoon2006}.
The calculations are started from models of uniformly rotating pre-main sequence stars, and are 
continued until at least core carbon depletion, and up to the formation of an iron core in most cases. 
Models above $\sim \ 50\mso$ may evolve until the onset of the pulsational pair instability (\citealt{2017ApJ...836..244W}). We used the nuclear network \texttt{approx21} included in MESA.

Convection is modelled with the standard mixing length theory \citep{1958ZA.....46..108B}, using $\alpha_{ML} = 1.5$. We adopt the Ledoux criterion for 
convective instability, and apply semiconvective mixing with an efficiency factor of $\alpha_{SC} = 0.01$ \citep{1991A&A...252..669L}.  Convective overshooting 
is applied by using a step function, with $\alpha_{ov} = 0.335$ following \cite{Brott2011}. 

We compute two sets of models, using different efficiency parameters for rotational mixing. In our first set
(the A\,Series),
we adopt the following parameters:
Our treatment of mixing includes Eddington-Sweet circulations, secular and dynamical instability, and 
the Goldreich-Schubert-Fricke instability, with an efficiency factor of $f_c = 1/30$ following \cite{Brott2011}. 
We employ the Tayler-Spruit dynamo to compute the internal magnetic field strength and the corresponding
transport of angular momentum as described in and \cite{HegerWoosleySpruit2005}.  
Time smoothing of the mixing coefficients is included to make our results consistent with those of \cite{Szecsi2015}.

In our second set of models (the B\,Series), we enhance the diffusion coefficient due to rotational mixing by a factor of ten.
While such high rotational mixing efficiency may lead to too strong mixing of nitrogen in slowly and
moderately rotating O\,stars \citep{Brott2011}, we use it here as a tool to produce rapidly rotating
progenitor models which are suitable to explain Type\,Ic supernovae. 
While it can not be excluded that at very high rotation rates,
non-linear effects lead to a stronger dependence of the
mixing efficiency on the rotation rate than at moderate or slow rotation,
we argue below that in any case, our major results are merely independent 
of how a rapidly rotating SN\,Ic progenitor model has actually been produced.

Mass loss rates for H rich (with a surface H mass fraction $X_S>0.7$) and for H poor models ($X_S<0.4$), were computed following \cite{Vink2001} and \cite{Hamann1995}, respectively, the latter multiplied by a factor of 1/10 to account for subsequent calibrations (cf. Fig.\,1 of \citealt{2005A&A...443..643Y}), and both using a metallicity scaling of $\dot M \sim (Z/Z_{\odot})^{0.85}$.
For intermediate surface H abundances we smoothly interpolate between the two, as in \cite{Marchant2016} and \cite{Yoon2006}.
We apply a mass loss enhancement based on the ratio of the rotation rate at the stellar surface to the critical rotation rate  
\begin{equation}
\Omega_{crit} = \sqrt{\frac{GM}{R^3}(1-\Gamma)},
\end{equation}
where $\Gamma = L/L_{Edd}$ is the ratio of the luminosity of the star to its Eddington luminosity; as
\begin{equation}\label{eq:mdot-rot}
\dot{M}(\Omega) = \dot{M}(0)\left(\frac{1}{1-\frac{\Omega}{\Omega_{\rm{crit}}}}\right)^{\xi},
\end{equation}
where $\xi = 0.43$ \citep{1986ApJ...311..701F,1993ApJ...409..429B}.
Models with  $\Omega / \Omega_{\rm{crit}}$ approaching unity lead to the divergence of $\dot{M}$ according to Eq. \ref{eq:mdot-rot}. To work around this, we set the maximum value for $\Omega / \Omega_{\rm{crit}}$ to 0.98, calculating the mass loss rate implicitly when this limit was exceeded, such that over-critical rotation and diverging mass loss rates are avoided. 
We note that the resulting mass loss rate is well defined and insensitive to the 
chosen threshold value \citep{Langer98}.

The details of our mass loss enhancement procedure are to a large part chosen to facilitate numerical
convergence, but do not affect our main results. Its practical purpose is to determine the amount of mass loss which is required for
the star to not exceed the limit of critical rotation.
As the radiation driven winds of our low metallicity models
are weak, this amount is 
determined simply by the initial conditions of the 
models, and the structural changes imposed by the
internal stellar evolution.

\section{Model properties} \label{sec:results}

The observed dependence of the stellar wind mass loss rate on metallicity \citep{2007A&A...473..603M} implies that metal poor stars lose little mass during their evolution. However, in rapid rotators, centrifugally driven mass loss may cause high mass loss rates
even at low metallicity during overall contraction stages that 
occurs between major nuclear burning stages \citep{Yoon2006,2008A&A...478..467E}.
Such events will heavily affect the circumstellar medium (CSM) structure of 
such stars \citep{vanMarle2008}, 
which may lead to different pre-SN configurations in different circumstances.

To investigate this, we perform evolutionary calculations for rapidly rotating
stars in the initial mass range from 5 to 100 $\rm{M}_{\odot}$, which undergo CHE.
Our Series\,A models follow closely the paths described by \citep{Yoon2006} and \cite{Szecsi2015},
and we describe them below for comparison. However, many of these models retain massive helium-rich envelopes
and might thus correspond to progenitors of Type\,Ib supernovae.
We therefore focus below on our Series\,B models, in which the rotational mixing remains strong enough
during core helium burning such that little helium remains. As we shall see, this does not
only affect the spectroscopic type of the potentially produced SNe, but it profoundly changes
the evolution of the internal angular momentum and of the CSM.

\subsection{Evolution of the internal chemical structure}\label{chem}

\begin{figure*}[ht!]
\plotone{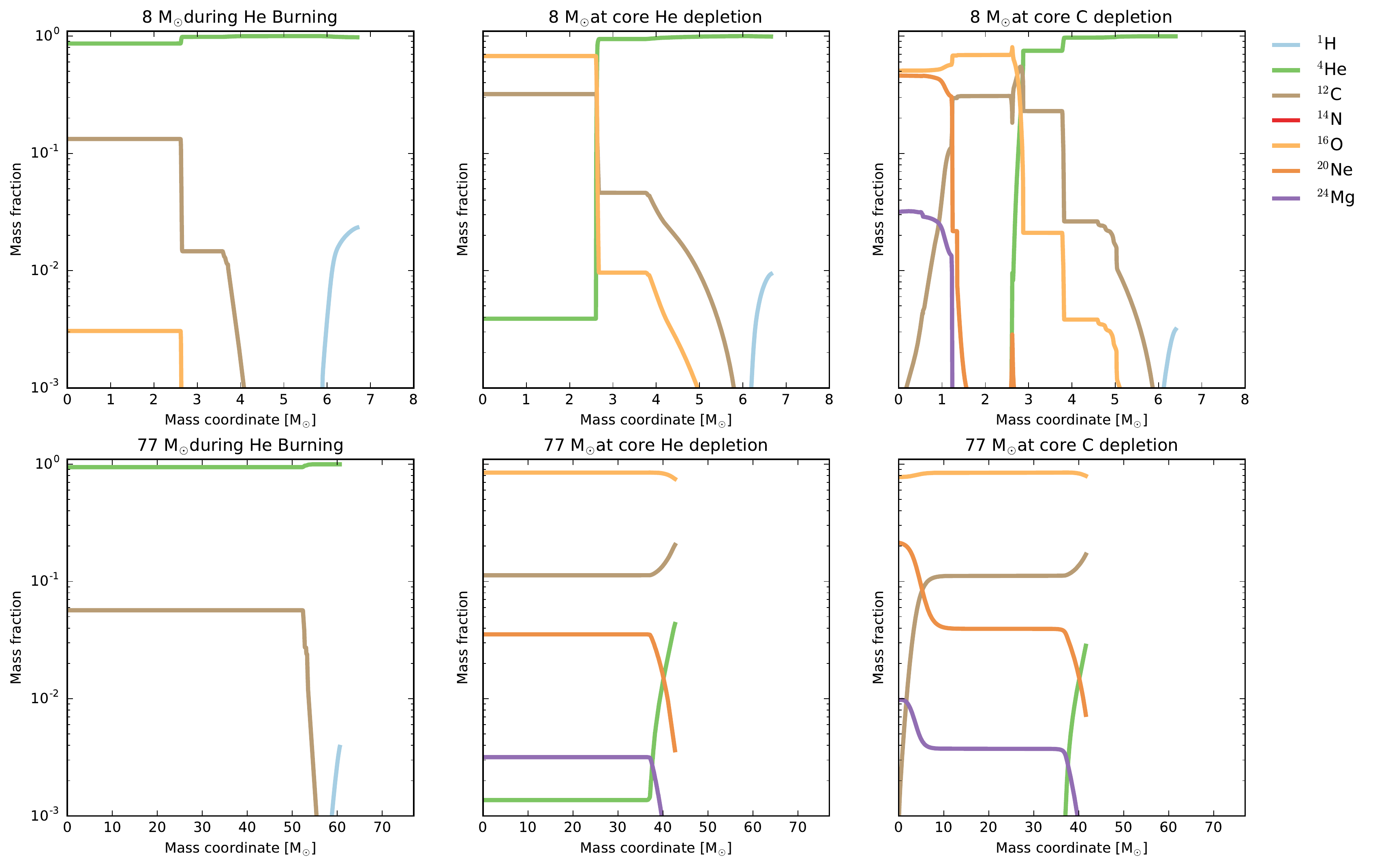}
\caption{Mass fractions of the major chemical elements as function of
the Langrangian mass coordinate in stellar models from two 
evolutionary sequences of the A\,Series at three different times, during core He burning, 
at He core depletion, and at C core depletion.}\label{fig:abundances}
\end{figure*}

\begin{figure*}[ht!]
\plotone{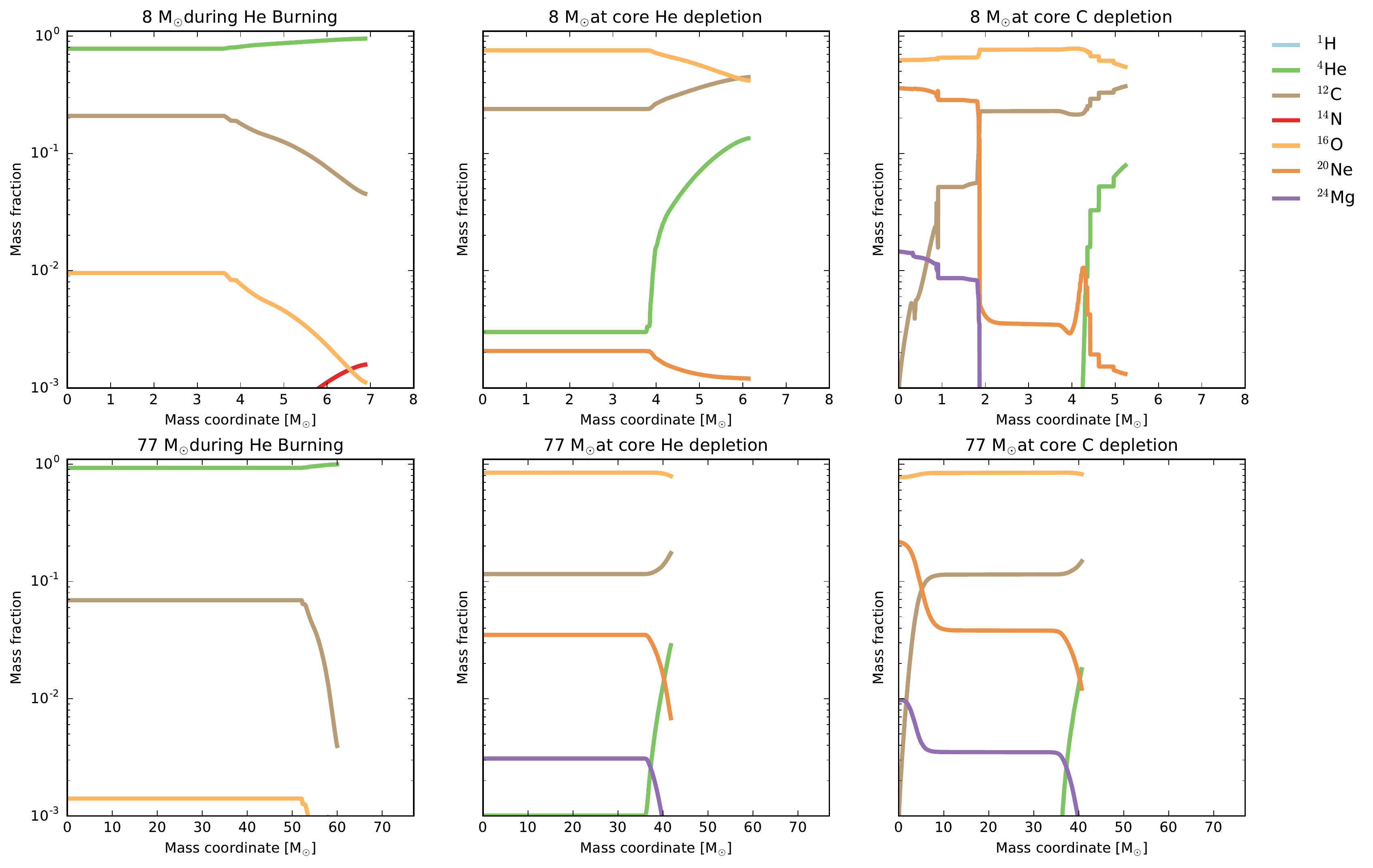}
\caption{As Figure\,1, but for the corresponding models from the B\,Series.}\label{fig:abundancesB}
\end{figure*}

As described in Sect. \ref{sec:method}, our initial models have a uniform composition, and they evolve 
quasi-chemically homogeneously throughout core H burning. As they transition to core He burning, 
they contract, and a significant rotationally enhanced mass loss occurs (see Sect. \ref{sub:structure}).
As illustrated in Figs. \ref{fig:abundances} and \ref{fig:abundancesB}, H is 
essentially absent at the pre-SN stage in all our models, regardless of their mass. 
This is due in part to the mixing that brings most of the H into the core, 
and also to the rotationally enhanced mass loss that removes the remaining H from the surface.

In our Series\,A models, the extent of the region where mixing is efficient, 
and the amount of mass lost during contraction determine whether a particular evolutionary 
sequence remains quasi-chemically homogeneous throughout He burning, or instead
develops a core-envelope structure in this phase.
Two models from Series\,A sequences shown in Fig.\,\ref{fig:abundances} illustrate the two different cases: 
models below $\sim 20 \ \rm{M}_{\odot}$ retain He-rich envelopes until after core He burning, 
and ignite a vigorous helium shell burning.
More massive models produce less massive helium envelopes, with helium nearly absent 
in the highest masses (cf., Tab.\,\ref{tab:data}). 
He-shell burning is weak or absent in these models, and has no significant effect.

Our Series\,B models retain little helium ($< 0.06\,\rm{M}_{\odot}$) regardless of the initial mass 
(see Fig.\,\ref{fig:abundancesB} and Tab.\,\ref{tab:data2}). These models do not develop a core-envelope 
structure and remain well mixed also during He burning. None of these models develops significant
He-shell burning.

\subsection{Core contraction and envelope expansion}\label{con}

\begin{figure}[ht!]
\plotone{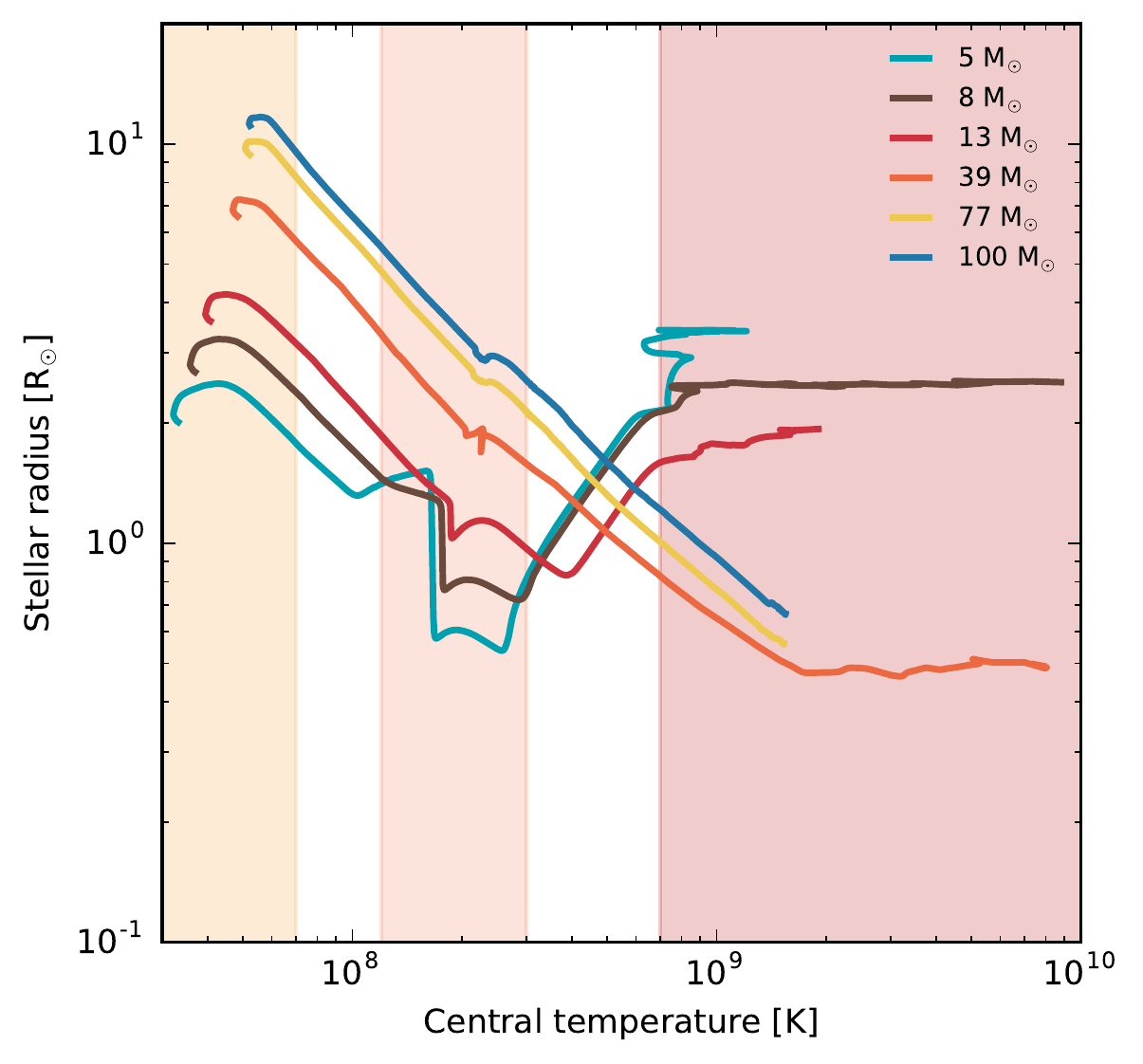}
\plotone{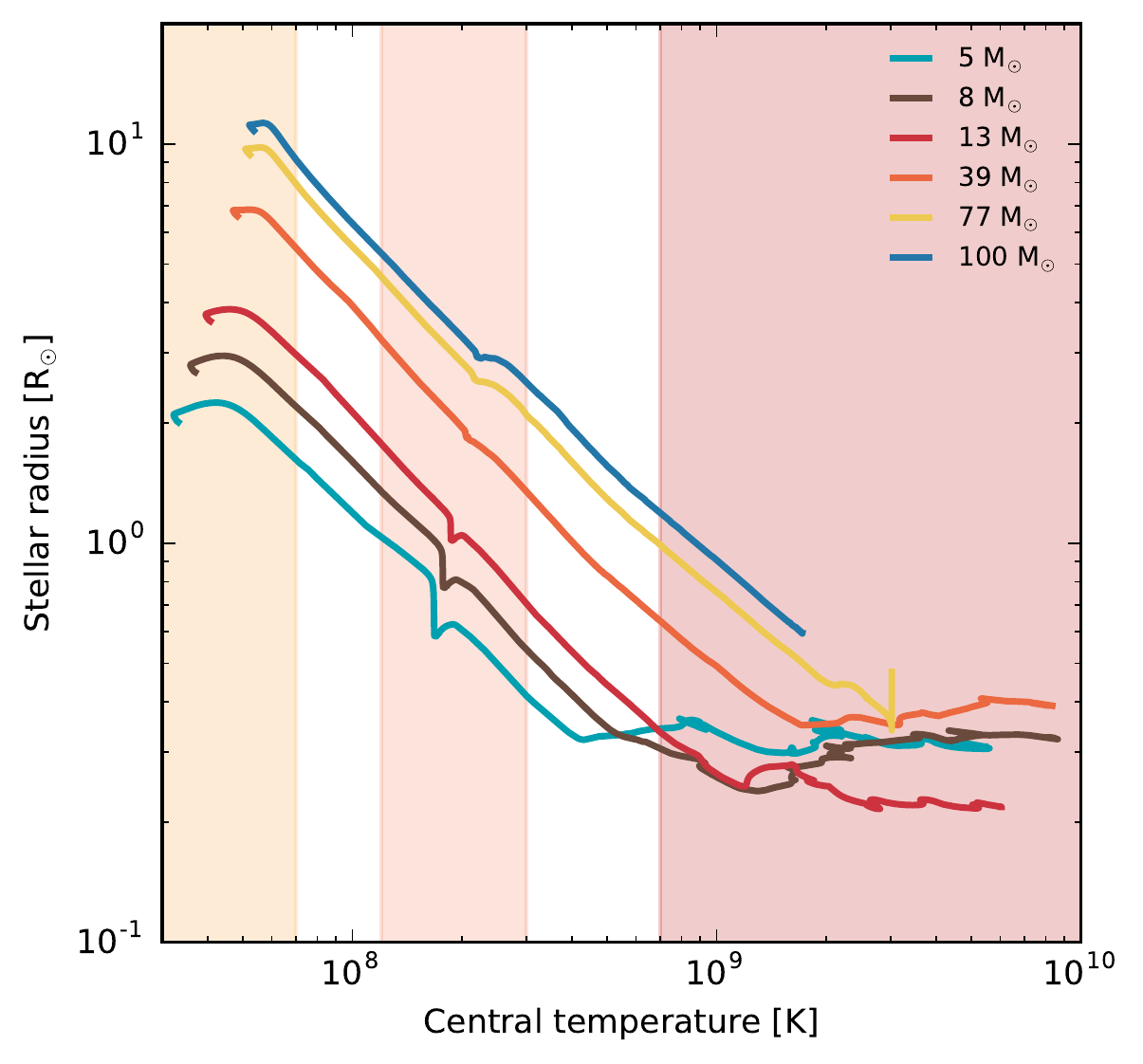}
\caption{Stellar radius as a function of central temperature for the evolutionary sequences of Series\,A (top) and
Series\,B (bottom). The temperature ranges of core H, He and heavy element burning are shaded.
}\label{fig:radii}
\end{figure}

As all massive star models, our models experience an overall contraction
immediately after core H exhaustion (Fig.\,\ref{fig:radii}). 
While in the stellar core, this contraction continues until core helium ignition,
the ignition of a powerful H shell source stops the contraction of the envelope and leads to its rapid expansion in stellar
models which follow ordinary evolution (cf. \citealt{Brott2011}). 

\begin{figure}[ht!]
\plotone{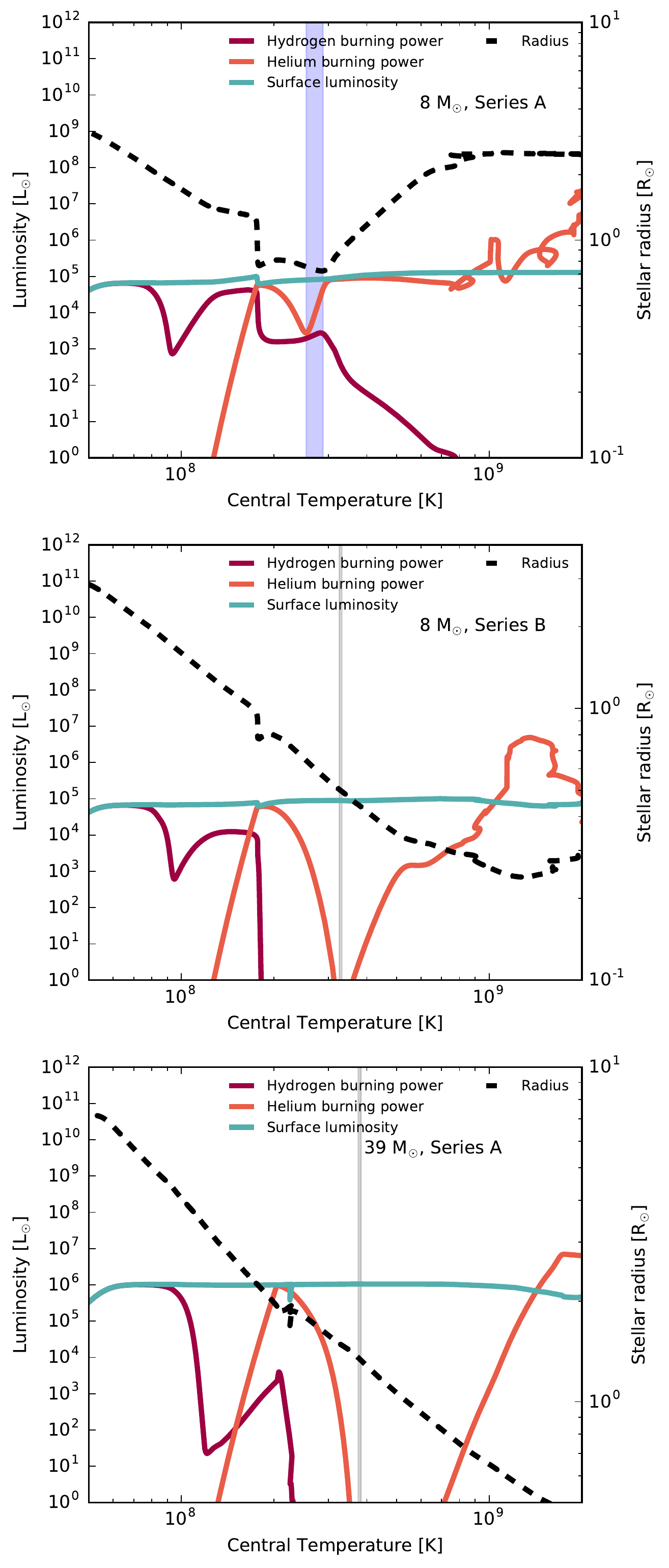}
\caption{Nuclear energy produced per second by H and helium burning, surface luminosity (solid lines, left axis), and stellar radius (dashed line, right axis), as a function of central temperature, for the 8 M$_{\odot}$ (top) and 39 M$_{\odot}$ sequences (bottom) from Series\,A. The blue highlighted region corresponds to the duration of one Kelvin-Helmholtz timescale beginning at He shell ignition in the top panel. Bottom panels show instead the temperature of shell ignition, but the remaining lifetime is shorter than the Kelvin-Helmholtz timescale at this point.}\label{fig:shells}
\end{figure}

In our models, which experience CHE, the latter effect is weak or absent.
Since little H is left at core H exhaustion in our massive models ($M > 20\mso$; see Sect. \ref{chem}), burning in their H shells is
weak, and they show no expansion at this stage.
Our lowest mass models of the A\,Series do experience a H-shell driven expansion (at $T_{\rm c}\simeq 10^8\,$K; see Fig.\,\ref{fig:shells}). 
However, most of their remaining H is quickly lost due to rotationally enhanced mass loss (Sect. \,\ref{sub:structure}) and
H-shell burning, such that their envelops contract soon thereafter. During core helium burning ($T_{\rm c}\simeq 10^8\,$K),
the radii of all the models correspond well to the mass-radius relation of helium stars \citep{1989A&A...210...93L}.

Most remarkably, Fig.\,\ref{fig:radii} shows a pronounced dichotomy in the final radius evolution of our models. Whereas the three lower mass Series\,A models strongly expand after core helium exhaustion, all other models continue to contract.

The expansion of the lower mass Series\,A models is driven by
the onset of strong helium shell burning. As shown in Fig.\,\ref{fig:shells} (top)
at the example of our 8\,M$_{\odot}$ model, the final radius increase in this
sequence starts about one thermal timescale after the onset of helium shell burning.

Fig.\,\ref{fig:shells} (middle) shows, again with our 8\,M$_{\odot}$ model as an example, why this expansion does not occur in our Series\,B models. The reason is that
they contain much less helium in their envelopes at this stage, such that
the helium shell source is much weaker or absent. Therefore, even the lower
mass models of Series\,B contract as a whole after core helium exhaustion.
The contracting models end their lives as very compact and extremely hot stars,
with $T_{\rm eff} \simeq 150\,000\dots 300\,000\,$K (cf. Tables \ref{tab:data2} and \ref{tab:data}).

Fig.\,\ref{fig:shells} also shows why the massive models of Series\,A do not expand after core
helium exhaustion. The amount of helium in the 39\,M$_{\odot}$ model at this stage is still
large ($\sim 4\,$M$_{\odot}$; cf., Tab.\,\ref{tab:data}), and as shown in the figure, the energy production 
in the helium shell source exceeds the stellar luminosity by far at some point. 
However, due to the copious neutrino emission in the stellar core, the remaining life time
of the model at this stage is significantly smaller than its Kelvin-Helmholtz timescale. 
It undergoes the post-core helium exhaustion overall contraction, 
and reaches core collapse before the helium shell burning can expand the envelope.

\begin{figure}[ht!]
\epsscale{0.95}
\plotone{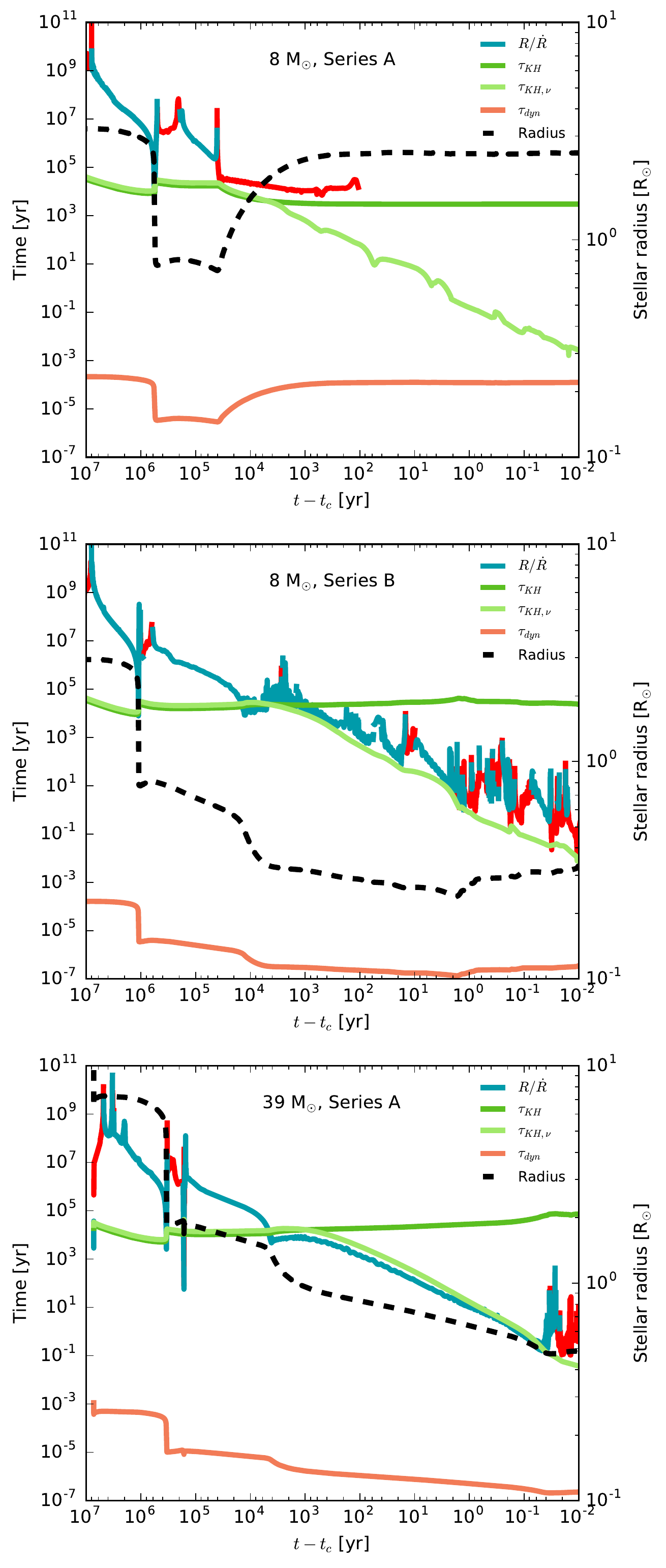}

  \caption{Time-scale of the radius evolution $R$/$\dot{R}$ (in blue for contraction, in red for expansion), Kelvin-Helmholtz time-scale $\tau_{KH}$ (Eq.\,\ref{eq:kha}), neutrino-mediated thermal time-scale $\tau_{KH, \nu}$ (Eq.\,\ref{eq:kh}), and dynamical time-scale $\tau_{dyn}$ as a function of the remaining time to core collapse, for the 8\,M$_{\odot}$ models
of Series\,A (top) and Series\,B (middle), and for the 39\,M$_{\odot}$ model of Series\,A (bottom). Parts of the $R$/$\dot{R}$ line
are omitted as the radius (black dashed line) becomes almost constant, and numerical noise dominates due to small 
changes of the radius.}\label{fig:timescales}
\end{figure}

From Fig.\,\ref{fig:timescales}, we can understand the timescale of the radius change of our models, by comparing
it to various other timescales. As expected, during core H and core helium burning, the stellar
radius changes on the nuclear timescale (not shown). This is 2 to 3 orders of magnitude larger than
the Kelvin-Helmholtz timescale
\begin{equation}\label{eq:kha}
\tau_{\rm{KH}} = \frac{GM^2}{R L},
\end{equation}
on which the stellar radius changes at core H exhaustion. 
However, those models which contract after core helium
exhaustion do so on a much shorter timescale. With the neutrino-mediated Kelvin-Helmholtz timescale defined as
\begin{equation}\label{eq:kh}
	\tau_{\rm{KH, \nu}} = \frac{GM^2}{R(L+L_{\nu})},
\end{equation}
where $L_{\nu}$ is the neutrino luminosity, Fig.\,\ref{fig:timescales} shows that the contraction occurs on this timescale.
Therefore, on one hand, these models, which are essentially just bare C/O-cores, contract to compensate for their
neutrino losses. 

On the other hand, the top panel of Fig.\,\ref{fig:timescales} shows that our 8\,M$_{\odot}$ Series\,A model
performs its final expansion on the ordinary Kelvin-Helmholtz timescale, even though its C/O-core also
contracts on the neutrino-mediated Kelvin-Helmholtz timescale. This demonstrates that only contraction can be
accelerated by neutrino losses, but evidently expansion can not, since the neutrinos 
do not contribute to the pressure inside the star. 

Therefore, even if the core contraction leads to the ignition of a vigorous
helium shell source (as it is the case in our $39\mso$ sequence; cf. Fig.\,\ref{fig:shells}), but the time to core collapse is
significantly shorter than the ordinary Kelvin-Helmholtz timescale, the star will not be able to expand any more.
Consequently, our $39\mso$\,model decreases its radius monotonically as function of time until it reaches a minimum, at which it stays until the end of our calculations (Fig.\,\ref{fig:radii}).

We note that during the pre-collapse evolution of ordinary red supergiants, the cores after core helium exhaustion must contract
on the neutrino-mediated Kelvin-Helmholtz timescale. As a consequence, in all of them a vigorous helium shell
burning will start. Whether or not this will give rise to a subsequent stellar expansion will again depend on the
mass of the star. For the highest masses, the core contraction timescale will be smaller than the stellar Kelvin-Helmholtz 
timescale, and no expansion will occur. For lower masses, in particular when core degeneracy slows down the core contraction,
the envelope has ample time to expand, and the stellar track will climb up the Hayashi line in the HR diagram
(see e.g. Fig.\,1 of \citealt{2012A&A...542A..29G}).

\begin{figure*}[ht!]
\plotone{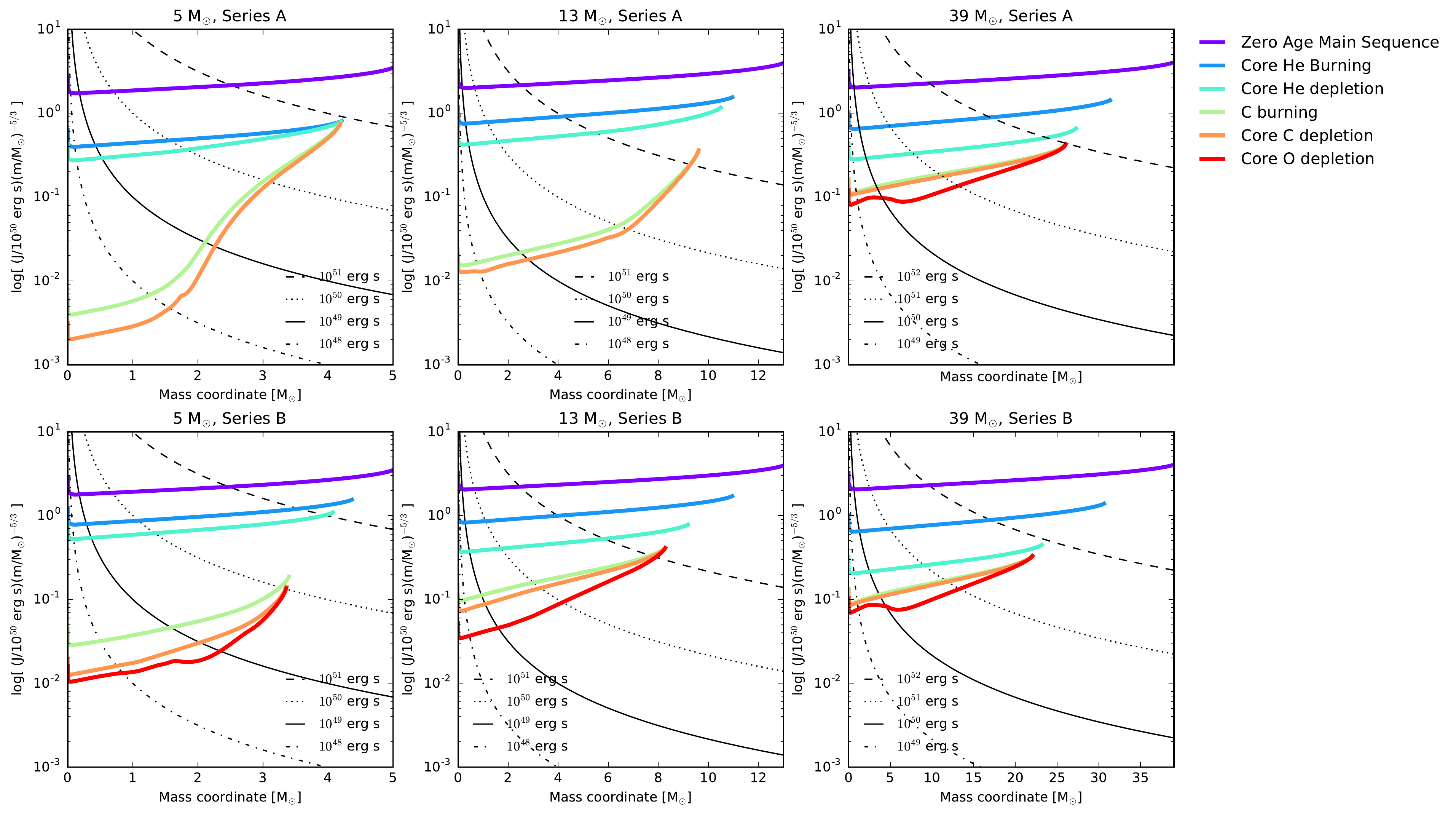}
\caption{Integrated angular momentum $J(m)$ divided by $m^{5/3}$ as a function of the model's mass coordinate, at main sequence (purple), start of core He burning (dark blue), core He depletion (light blue), during C burning (green), at C core depletion (orange) and at core O depletion (red), for Series\,A evolutionary sequences. Contours (labeled individually for each panel) represent lines of constant $J$.}\label{fig:angularmomentum}
\end{figure*}

\subsection{Evolution of the internal angular momentum distribution}\label{sec:angu}

The distribution of angular momentum in our evolutionary models initially corresponds to uniform rotation.
Its evolution is calculated by considering the combined effects of hydrodynamic and magnetic angular momentum transport, 
and angular momentum loss due to mass loss. In particular, the Spruit-Taylor mechanism, which has been invoked to explain the 
observed slow rotation rates of white dwarfs \citep{Suijs2008} and young pulsars (\citealt{HegerWoosleySpruit2005}),
can efficiently transport angular momentum from the core into the envelope of our models.

Figure \ref{fig:angularmomentum} shows the distribution of angular momentum in the stellar interior in several stages of evolution, for selected models. We show this following \cite{Suijs2008}, by plotting the total angular momentum enclosed inside a given mass coordinate $m$, divided by $m^{5/3}$; namely $(m)^{-5/3} J(m) = (m)^{-5/3}\int_0^{m}j(m')\text{d}m'$, where $j(m')$ is the specific angular momentum at mass coordinate $m'$. A uniformly rotating sphere of constant density will have flat a profile in this diagram.

Figure \ref{fig:angularmomentum} shows that in the lower mass models (5$\dots$20\,M$_{\odot}$), a significant
amount of angular momentum transport occurs after core helium exhaustion. This increases the specific angular momentum
in the envelope, and therefore, at a given radius, the surface rotation of the models. In the lower mass models, 
this effect leads to an enhancement of the rotationally induced mass loss despite their expansion (see Sect.\,\ref{sub:structure}).

\begin{figure}[ht!]
\plotone{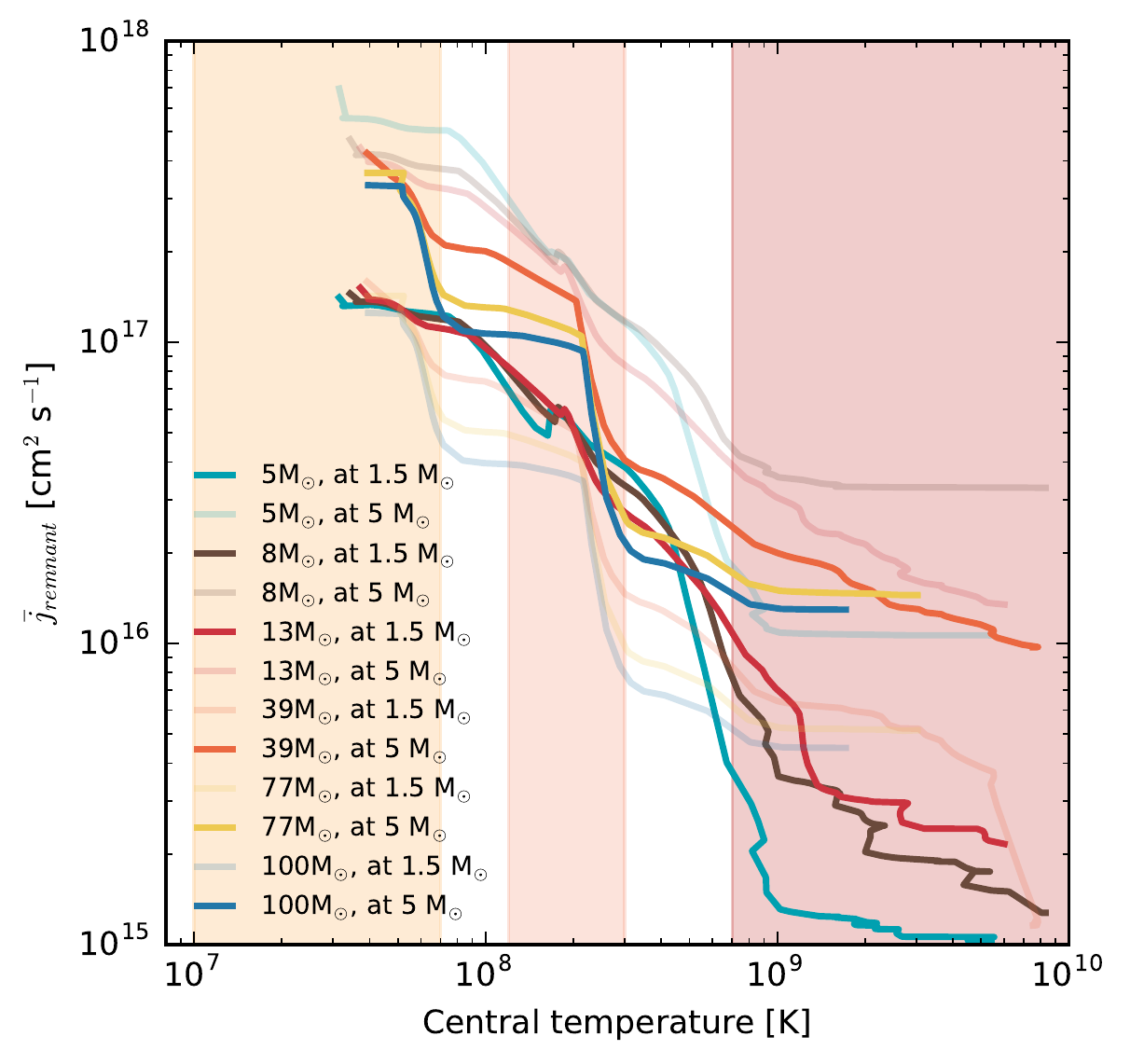}
\caption{Average specific angular momentum of the innermost 1.5$\,\mso$ for models with $M < 20\,\mso$, 
and of the innermost 5$\,\mso$ for models with $M \geq 20 \mso$ of our Series\,B models, as a function of central temperature in solid lines. The average specific angular momentum of the innermost 5$\,\mso$ for models with $M < 20\,\mso$, 
and of the innermost 1.5$\,\mso$ for models with $M \geq 20 \mso$ are shown in transparent lines, for comparison.
}\label{fig:jvsT}
\end{figure}

In the more massive models, the short core contraction timescale does not allow for significant 
angular momentum transport after core helium exhaustion. The angular momentum distribution remains 
almost frozen-in. Therefore, even though the wind induced amount of mass lost by the massive stars is larger,
their final specific core angular momentum is larger than that of the lower mass models. This is illustrated in Fig. \ref{fig:jvsT}, which shows the time evolution of the average specific angular momentum of the core.

In Tabs. \ref{tab:data2} and\,\ref{tab:data}, we provide the final core specific angular momentum of our models,
and estimate the initial spin period of a potential neutron star emerging from the collapsing core. 
The obtained results for Series\,A agree qualitatively with the lowest metallicity models of \cite{Yoon2006}.

The cores of the low mass models of Series\,A appear to rotate too slowly to apply either,
the magnetar or the collapsar scenario to them. Due to the core-envelope structure established 
during core helium burning, the expanding helium envelope produces a strong braking of the 
core rotation after core-helium exhaustion. The cores of the massive models retain sufficient
angular momentum to qualify for the collapsar scenario, but only the most massive models
die with less than one solar mass of helium left in their envelopes.

Our Series\,B models are all essentially helium-free at the time of collapse, and all
of them have average specific core angular momenta of more than $10^{15}$\,cm$^2$\,s$^{-1}$.
Certainly the 5$\,\mso$ and 8$\,\mso$ sequences, with final masses of 3.4$\,\mso$ and 5.2$\,\mso$, 
and possibly also the 13$\,\mso$ sequences ending up with 8.3$\,\mso$,
can be expected to form neutron stars. Assuming angular momentum conservation during core collapse,
their initial spin periods would be expected in the range $5\dots 2\,$ms 
(cf. Tab.\,\ref{tab:data2}), suitable for the scenario of magnetar driven superluminous supernovae \citep{2015MNRAS.454.3311M}.

Assuming that the more massive Series\,B models would form black holes of at least 5$\,\mso$
\citep{2017MNRAS.469L..43O},
their average specific angular momentum would be $10^{16}$\,cm$^2$\,s$^{-1}$ or more,
implying they would be progenitor candidates for lGRBs within the collapsar scenario
\citep{Woosley1993} (see Sect.\,\ref{sec:consequences} for a detailed discussion).

\subsection{Mass loss history}\label{sub:structure}

\begin{figure}[ht!]
\plotone{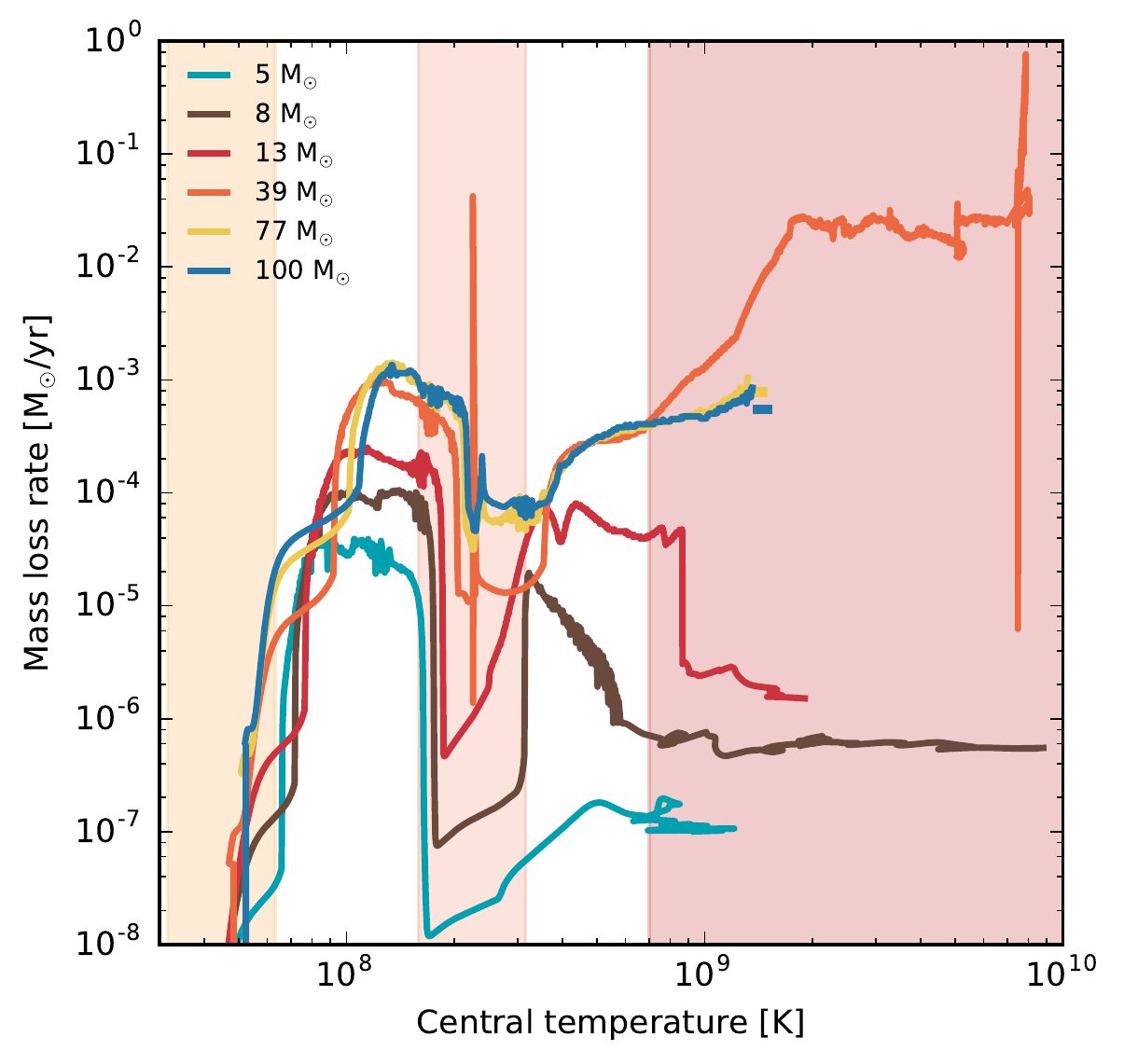}
\plotone{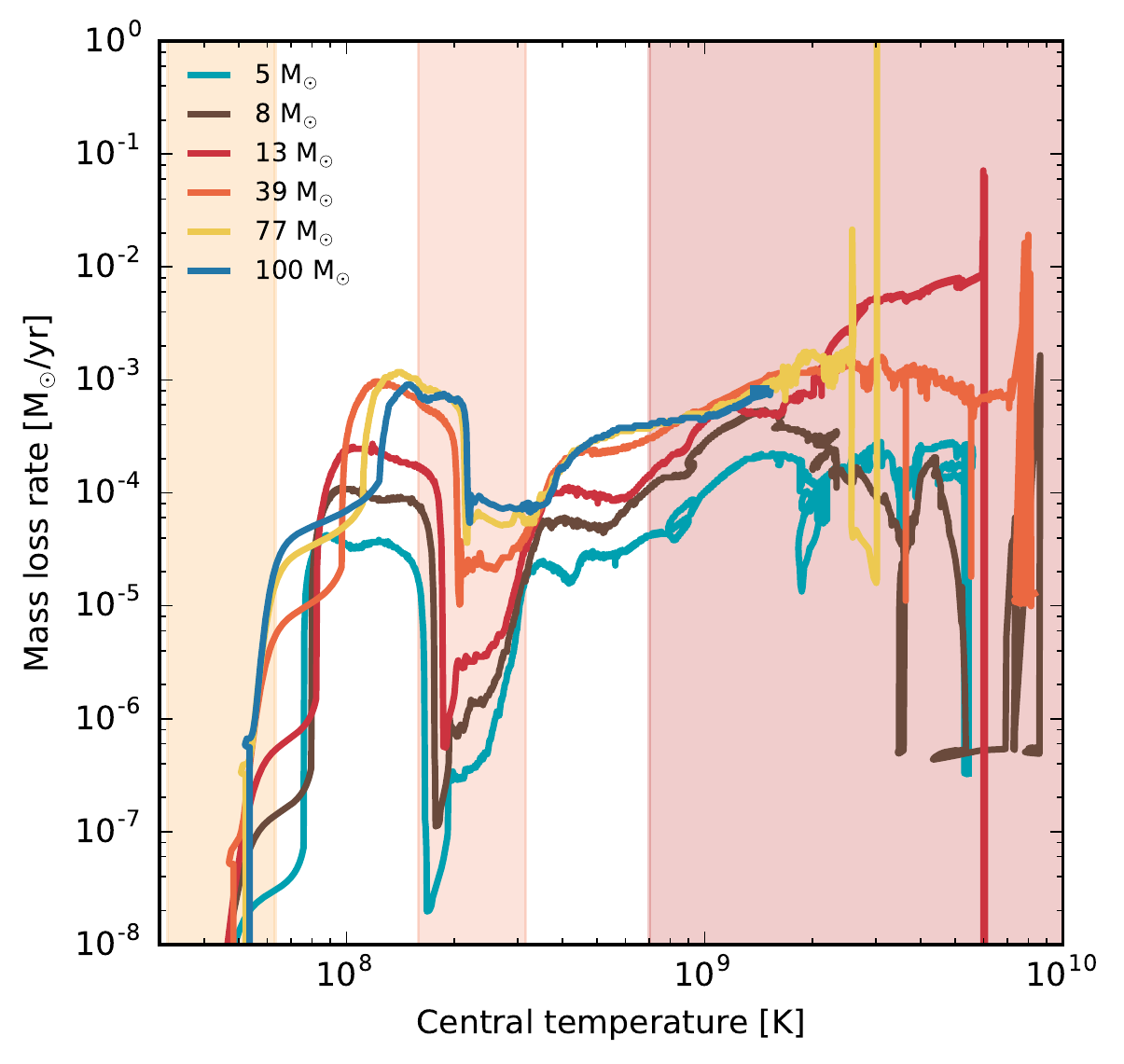}
\caption{Mass loss rate as a function of central temperature for the evolutionary of Series\,A (top) and
Series\,B (bottom). The temperature ranges with H, He and heavy element burning are highlighted.}
\label{fig:mdot}
\end{figure}

As the adopted metallicity of our models is very small, radiation driven stellar winds
are expected to be weak, and we do not discuss them here. Instead, as shown in Fig.\,\ref{fig:mdot}, 
the mass loss rates of our models can become large in the contraction phases following
core H and core helium burning. The reason is that the spin-up of the surface layers which
accompanies the contraction would lead to over-critical rotation if the mass loss would not be enhanced.
In this situation, even though the mass loss rate is formally computed by\,Eq. \ref{eq:mdot-rot}, and by MESA's implicit mass loss routine in the case of near-critical rotation, it is not affected
by the stellar wind mass loss rate, but rather determined by the amount of angular momentum which is required
to be lost from the star for avoiding over-critical rotation \citep{Langer98}. As such, the mass loss rate is
an inherent prediction of our model calculations.
Whether or not the mass lost in this way can be transported to infinity, and if so which force could do this,
depends on the achieved mass loss rate and will be discussed below. In our models, we argue that
once mass is lost from the star it will not fall back, since the star is already very close to critical rotation.

Focusing on the mass loss rate increase after core helium exhaustion, we find it to be driven by 
two mechanisms. Most important, the models which undergo an overall contraction at this stage 
spin up, which leads them quickly to approach critical rotation. Strong mass loss is then required to
avoid over-critical rotation. However, as shown in Fig.\,\ref{fig:mdot}, also the lower mass models
of Series\,A, which expand after core helium exhaustion, show a partly significant increase in the mass loss rate.
This is driven by the angular momentum transport in the lower mass models
(cf., Sect.\,\ref{sec:angu}), which can be
more significant than the spin-down effect due to the envelope expansion.

The overall post core-helium exhaustion mass loss of our model sequences  (see Tabs. \ref{tab:data2} and \ref{tab:data})
is of the order of one solar mass, except for the 5\,M$_{\odot}$ and the 8\,M$_{\odot}$ models
of Series\,A. Whereas the achieved mass loss rates in the more massive models are typically larger,
their post-core helium burning life times are smaller. Also, in the lower mass models of Series\,B,
both mass loss rate enhancement mechanisms work simultaneously, such that the corresponding
5\,M$_{\odot}$ model achieves the largest relative amount of mass being lost at this stage.

From the constant slope of the line for the contraction timescale in Fig.\,\ref{fig:timescales}, it follows that
the contraction accelerates according to a power law, and so does the mass loss. While the mass loss
rates remain mostly around 10$^{-3}$\,M$_{\odot}$yr$^{-1}$, they do approach 10$^{-2}$\,M$_{\odot}$yr$^{-1}$
in the 39\,M$_{\odot}$ models.

It is uncertain whether the mass shed by our models during the later phases can be accelerated to large radii, since it is not due to radiatively accelerated winds, but dominated by the centrifugal force, which drops quickly with distance to the star. At the highest
mass loss rates, assuming a wind velocity of only 500\,km/s,
the so-called photon-tiring limit 
\citep{1996A&A...315..421H, 1997ASPC..120..121O, 1998A&A...329..190G}
is violated. I.e., such winds would violate energy conservation.

\begin{figure}[ht!]
\plotone{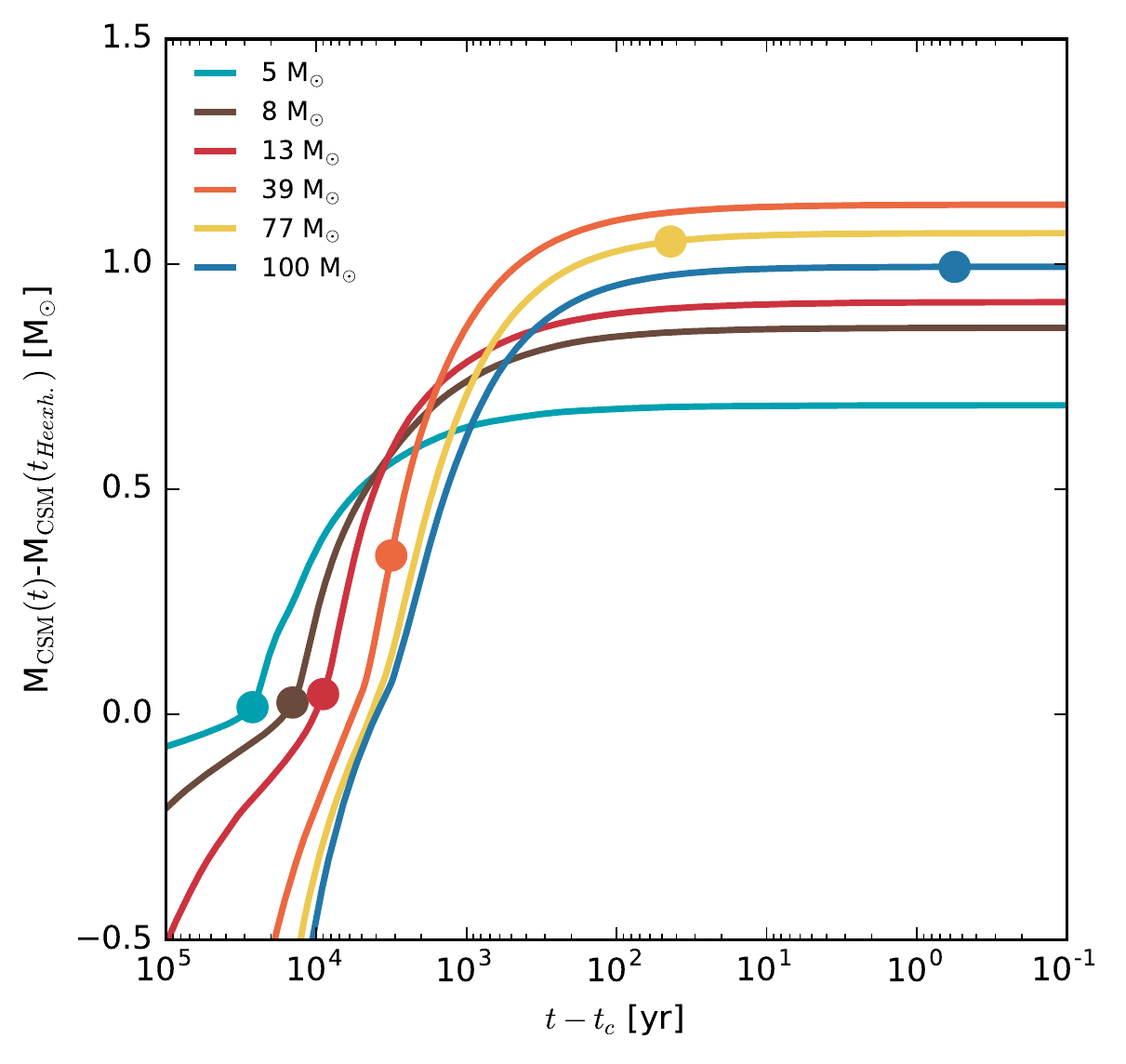}
\caption{Time integrated mass loss rate with the end of core
helium burning as zero point in time, as function of time,
for six models of our Series\,B (bottom). Dots indicate where the mass loss rate becomes 10 times higher than that allowed by the momentum limit.}
\label{fig:mdlim}
\end{figure}

The momentum limit of radiation driven winds is met at much smaller 
mass loss rates. While in Wolf-Rayet stars, multiple scattering
of photons allows this limit to be exceeded by a factor of a few
\citep{1994A&A...289..505S}, the winds may break down for much higher values.
In Fig.\,\ref{fig:mdlim} shows when in time the rate of the centrifugally driven mass loss exceeds the momentum limit by a factor of ten. 
We see that in particular for the lower mass sequences, the concerns
the bulk of the mass which is lost after core helium exhaustion.

We conclude that mass lost by the stars after core helium exhaustion is likely to remain nearby the star.
In addition, the outflows may be concentrated towards the equatorial plane of the star,
which leads to higher local mass densities compared to the case of isotropic mass loss \citep{vanMarle2008}.
On the other hand, as the spin-up accelerates with time until the collapse of the iron core,
and the mass loss rate increases with time correspondingly, any of the lost material is unlikely
to be re-accreted by the star before the time of the SN.
The chance to push the late mass loss to larger radii is largest
for our highest mass models. However, as we shall argue below, their
CSM at the time of their iron core collapse may be dominated by
massive shells ejected in the course of their pulsational pair instability.

\subsection{Magnetic fields}\label{sec:bfields}

The rotation profile of our evolutionary sequences remains close to uniform during the main sequence due to the torques induced by dynamo-generated magnetic fields \citep{2002A&A...381..923S,HegerWoosleySpruit2005}.  This mechanism, as discussed in Sect.\,\ref{sec:angu}, is the main component transporting angular momentum in our evolutionary sequences. However, it also has the effect of generating strong magnetic fields in the stellar interior, especially in the late phases of stellar evolution.

\begin{figure}[ht!]
\plotone{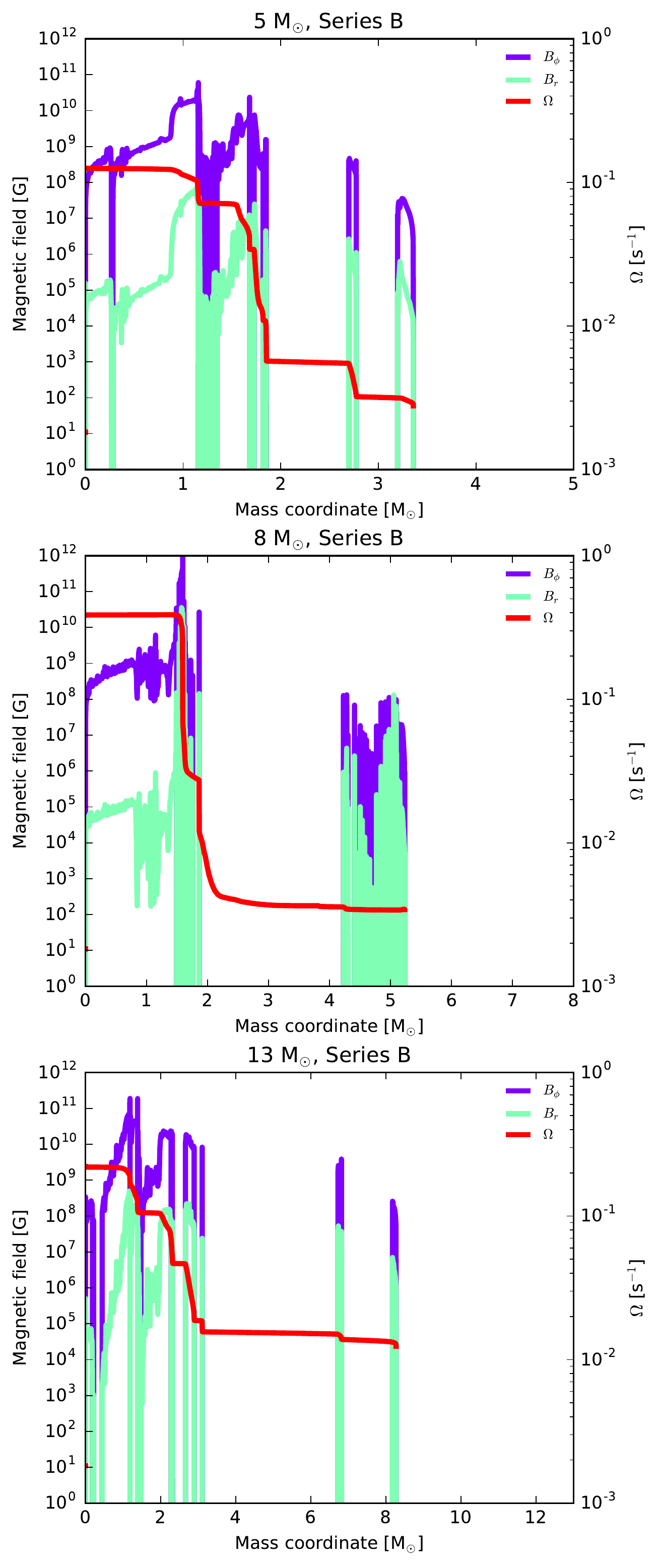}
\caption{Polar and radial magnetic fields at the pre-SN stage for 3 different evolutionary sequences from Series\,B, along with their rotational frequency.}\label{fig:magneticfields}
\end{figure}

Our pre-SN models, in particular, have strong magnetic fields, as illustrated in Fig.\,\ref{fig:magneticfields} for selected Series\,B models. As discussed in Sect.\,\ref{chem}, chemical homogeneity is lost at the start of C burning. At this time the rotation profile becomes increasingly stratified, allowing for strong shear layers which promote the formation of strong magnetic fields. The fields in our pre-SN models in Fig.\,\ref{fig:magneticfields} are of the order of $10^{10}\dots 10^{11}\,$G for the toroidal component and $10^{8}$ G for the radial component near the interface between the iron core and the Si burning shell. This is considerably larger than the $5 \times 10^{9}$ G field found in the 15 $\mso$ model of \cite{HegerWoosleySpruit2005}.
Assuming that the magnetic flux is preserved during core collapse, and that the produced neutron stars have 1.5 $\mso$ and a radius of 15\,km, leads to an enhancement of the B-field by a factor of $10^4$, i.e., 
to fields of order $10^{14}\dots 10^{15}$\,G after the collapse. While flux freezing is certainly an
oversimplification, according to \cite{2016MNRAS.460.3316R}), the MRI may enhance the B-field by up to a factor 
of\,10 during the iron core collapse. Thus, while we can not reliably predict the post-collapse B-fields
for our models, it appears plausible to assume that they do obtain magnetar strenghts.
 
Our most massive models are evolved to the onset of the pair instability before oxygen ignition.
At this stage, their toroidal core magnetic fields are of the order of $10^7\dots 10^8\,$G, which is
comparable to the field values of our lower mass models at the same stage of evolution.
According to \cite{2017ApJ...836..244W}, they can be expected to undergo a pulsational pair instability,
which will ultimately end with the formation of an iron core, iron core collapse, and black hole formation.
At the onset of the pair instability, the specific angular momentum of their cores exceeds 
$10^{16}\,$cm$^{2}$s$^{-1}$. 
Assuming that the cores of these models do not suffer strong angular momentum
loss during the pair-instability collapse and the further evolution towards iron core collapse
would imply that magnetic fields  would play a similar role in the final stages
of these models as they do for our lower mass models. However, the amount of core spin-down during the pulsational pair-instability process is yet unknown, and needs to be computed
from corresponding models which include rotation.

The core specific angular momentum in our Series\,B models is an increasing function of mass, and therefore 
also the core magnetic field strength is expected to increase with mass. Our lowest mass models will
form neutron stars, and the estimates above lead to magnetar strength B-fields.
Since the transition mass between neutron star and black hole formation in the case of magneto-rotational collapse,
and also the maximum neutron star mass in this case, is not well known (see \citealt{2017MNRAS.469L..43O}), our models
contain the possibility that the largest masses which lead neutron stars reach a core specific angular momentum of
$\sim 10^{16}\,$cm$^{2}$s$^{-1}$ and could thus produce magnetar driven GRBs \citep{2015MNRAS.454.3311M}.
Above this transition mass, black holes and collapsar driven GRBs are expected, where magnetic fields 
would still be likely to strongly affect the accretion flow to the black hole and the formation of the GRB jet.
As argued above, this may remain true even for our most massive models which undergo pulsational pair instability.

\begin{deluxetable*}{c|cccccc}
\tablecaption{Initial and final parameters of the evolutionary sequences in Series\,B.\label{tab:data2}}
\tablecolumns{7}
\tablewidth{0pt}
\tablehead{
\colhead{Initial mass [M$_{\odot}$]} & \colhead{5}  & \colhead{8}  & \colhead{13}  & \colhead{39}  & \colhead{77}  & \colhead{100} 
}
\setlength\extrarowheight{-1.6pt}
\startdata
Initial $\Omega / \Omega_{\rm{crit}}$ & 0.89 &  0.84 &  0.79 &  0.69 &  0.67 &  0.67 \\
CO core mass [M$_{\odot}$]  \tablenotemark{a} & 4.08 & 6.13 & 9.18 & 23.17 & 41.75 & 52.47 \\
Final mass [M$_{\odot}$] & 3.36 & 5.23 & 8.27 & 22.05 & 40.68 & 51.47 \\
$\Delta \rm{M}_{\rm H \rightarrow He}$ [M$_{\odot}$] \tablenotemark{b} & 0.36 & 0.55 & 1.04 & 4.58 & 13.58 & 20.41 \\
$\Delta \rm{M}_{\rm He \rightarrow final}   $[M$_{\odot}$]   \tablenotemark{c}   & $ 0.68 $  & $ 0.86 $& $ 0.92 $  & $ 1.13 $  & $ 1.07 $  & $ 0.99 $ \\
$f_{\rm M}$ & $ 0.27 $  & $ 0.19 $& $ 0.12 $  & $ 0.06 $  & $ 0.20 $  & $ 0.62 $ \\
Final Radius [cm] & $ 2.15 \times 10^{10} $  & $ 2.27 \times 10^{10} $& $ 1.52 \times 10^{10} $  & $ 2.72 \times 10^{10} $  & $ 3.31 \times 10^{10} $  & $ 4.14 \times 10^{10} $ \\
Final T$_{\rm{eff}}$ [K] & $ 1.63 \times 10^{5} $  & $ 1.72 \times 10^{5} $& $ 1.16 \times 10^{5} $  & $ 2.68 \times 10^{5} $  & $ 3.63 \times 10^{5} $  & $ 2.96 \times 10^{5} $ \\
Final H mass [M$_{\odot}$] & $ 0$  & $ 0 $& $ 0 $  & $0$  & $ 0$  & $ 0 $ \\
Final He mass [M$_{\odot}$] & $ 0.04 $ & $ 0.05  $ & $ 0.03 $ & $ 0.06 $ & $ 0.03  $ & $ 0.04  $ \\
Final Y$_{\rm{surf}}$ & $ 0.10 $ & $ 0.08$ & $ 0.07  $ & $ 0.03  $ & $ 0.02  $ & $ 0.02 $ \\
Radius at $\tau$ = 1 [cm] & $ 2.78 \times 10^{12} $  & $ 1.98 \times 10^{12} $& $ 3.32 \times 10^{13} $  & $ 2.16 \times 10^{13} $  & $ 1.67 \times 10^{13} $  & $ 1.33 \times 10^{13} $ \\
$\bar{j}_{\rm 1.5 \mso}$ [cm$^2$ s$^{-1}$] & $ 1.03 \times 10^{15} $  & $ 1.27 \times 10^{15} $& $ 2.16 \times 10^{15} $  & $ 1.15 \times 10^{15} $  & $ 5.15 \times 10^{15} $  & $ 4.50 \times 10^{15} $ \\
$\bar{j}_{\rm  2 \mso}$ [cm$^2$ s$^{-1}$]  & $ 1.38 \times 10^{15} $ & $ 1.92 \times 10^{15} $ & $ 3.17 \times 10^{15} $ & $ 2.38 \times 10^{15} $ & $6.35 \times 10^{15} $ & $  5.76 \times 10^{15} $ \\
$\bar{j}_{\rm  5 \mso}$ [cm$^2$ s$^{-1}$]  &  -- & $3.25 \times 10^{16}$ & $ 1.28 \times 10^{16} $ & $ 9.36 \times 10^{15} $ & $  1.29 \times 10^{16}$ & $ 1.16 \times 10^{16} $ \\
$P_{\rm rot}$ (1.5 $\mso$ NS) [ms] \tablenotemark{d} & 4.66 & 3.76 &  2.22 &  4.15 &  0.93 &  1.06 \\
$E_{\rm rot}$ (1.5 $\mso$ NS) [erg] \tablenotemark{d}&$ 2.07 \times 10^{51} $ & $ 3.18 \times 10^{51} $ &  $ 9.11 \times 10^{51} $ &  $ 2.60 \times 10^{51} $ &  $ 5.18 \times 10^{52} $ &  $ 3.96 \times 10^{52} $\\
Final fate &  CCSN  & CCSN &  CCSN  & CCSN   &  PPISN   &  PPISN  \\ 
\enddata
\tablenotetext{a}{At core He depletion}
\tablenotetext{b}{Mass lost from between H and He burning, defined as $\Delta \rm{M}_{\rm H \rightarrow He} = \int_{T8_{\rm c} > 0.5}^{T8_{\rm c} < 1.2} \dot M \, dt$}
\tablenotetext{c}{Mass lost from He core depletion to the end of the simulation}
\tablenotetext{d}{Assuming the remnant is a 1.5 $\mso$ neutron star, with 15 km radius, and with a moment of inertia of $2.27\times 10^{45}$ g cm$^2$, computed using Eq. 19 from \cite{2008ApJ...685..390W}}

\end{deluxetable*}
\section{Observable consequences} \label{sec:consequences}

In the previous section we have seen that our Series\,B models show some general properties:
They are all H-free and very helium-poor and thus qualify as progenitor models for
Type\,Ic supernovae, they all develop rapidly rotating cores and generate strong
magnetic fields before collapse, and they all suffer a rotationally induced mass loss of about $1\mso$ 
briefly before they produce a supernova. Here,
we will consider potential observational consequences of 
these properties during the supernova explosion. For this, we first discuss some general effects of
the expected pre-SN CSM on the shock break-out and supernova light curve. After this,
we discuss the more specific types of explosive events expected from our models as function of
their mass.

\subsection{Shock break-out}

The first electromagnetic signals from SNe are released by the shock break-out, the moment when the shock reaches near the stellar surface and photons start to be released from the shock front \citep[e.g.,][]{Colgate1974}. If an exploding star is surrounded by a dense CSM, the shock break-out signals can be strongly affected by it \citep{Ofek2010}.

The expected durations of the shock break-out from compact Wolf-Rayet progenitors are of the order of 1~sec. 
However, if they are surrounded by a dense CSM, photons released at the shock break-out are scattered 
in the dense CSM and the duration of the shock break-out can be extended 
to $\sim 100$~sec or more \citep[e.g.,][]{Balberg2011,Svirski2014}. 

A shock break-out from an exploding Wolf-Rayet star was observed in Type~Ib SN~2008D and 
its duration was indeed about 400~sec \citep{Soderberg2008}. Although the progenitor has helium in this case, 
our progenitors contain dense enough CSM to have similar extended shock break-out signals to those found in SN~2008D. 
However, the dense CSM is constrained to exist only within about $10^{15}~\mathrm{cm}$ 
in the case of SN~2008D \citep{Soderberg2008} and the dense CSM needs to exist only near 
the progenitor (see \citealt{Moriya2015} for a different idea). In addition, we note that 
this simple picture is based on a spherically symmetric model. 

For our models, the CSM at the time of the SN is expected to be very aspherical. 
If we assume that in the polar direction, along the rotation axis of the star, ordinary
radiation driven wind mass loss has dominated, shock break-out could occur along this
direction on the timescales mentioned above.

On the other hand, as argued in Sect.\,\ref{sub:structure}, the bulk of the centrifugally expelled mass shortly before the
supernova explosion might remain in a disk-like configuration embedding the star. In this case, the
SN shock may remain embedded for some time, and the early light curve of the SN could be strongly affected
by the SN-CSM interaction.

\subsection{CSM interaction}
 
When the SN ejecta collides with a dense CSM, the collision efficiently converts the ejecta kinetic 
energy to radiation. This extra radiation alters both, light curve and spectral properties of SNe (e.g. \citealt{2016ApJ...829...17S}). 
An estimate of the radiation energy produced by the interaction is obtained by 
assuming an inelastic collision of the ejecta with mass $M_{\rm ej}$ and velocity $\rm{v}_{\rm ej}$
with a shell with a mass of $M_{\rm sh}$ and velocity $\rm{v}_{\rm sh}$.
Momentum conservation results in a fraction of kinetic energy that is removed from the ejecta as
\begin{equation}\label{eq:deltaE}
\Delta E = E_{\rm kin,ej}  f_{\rm M} f_{\rm v}^2
\end{equation}
where $E_{\rm kin,ej} = M_{\rm ej} \rm{v}_{\rm ej}^2 / 2$ is the kinetic energy of the supernova,
and
\begin{equation}\label{eq:fm}
f_{\rm M} = {{M_{\rm sh}} \over {M_{\rm ej}+M_{\rm sh}}},
\end{equation}
and 
\begin{equation}\label{eq:fs}
f_{\rm v} = {\rm{v}_{\rm ej} -\rm{v}_{\rm sh} \over \rm{v}_{\rm ej}}
\end{equation}
account for the relative mass and velocity difference between shell and ejecta. 

\begin{figure*}[ht!]
\plotone{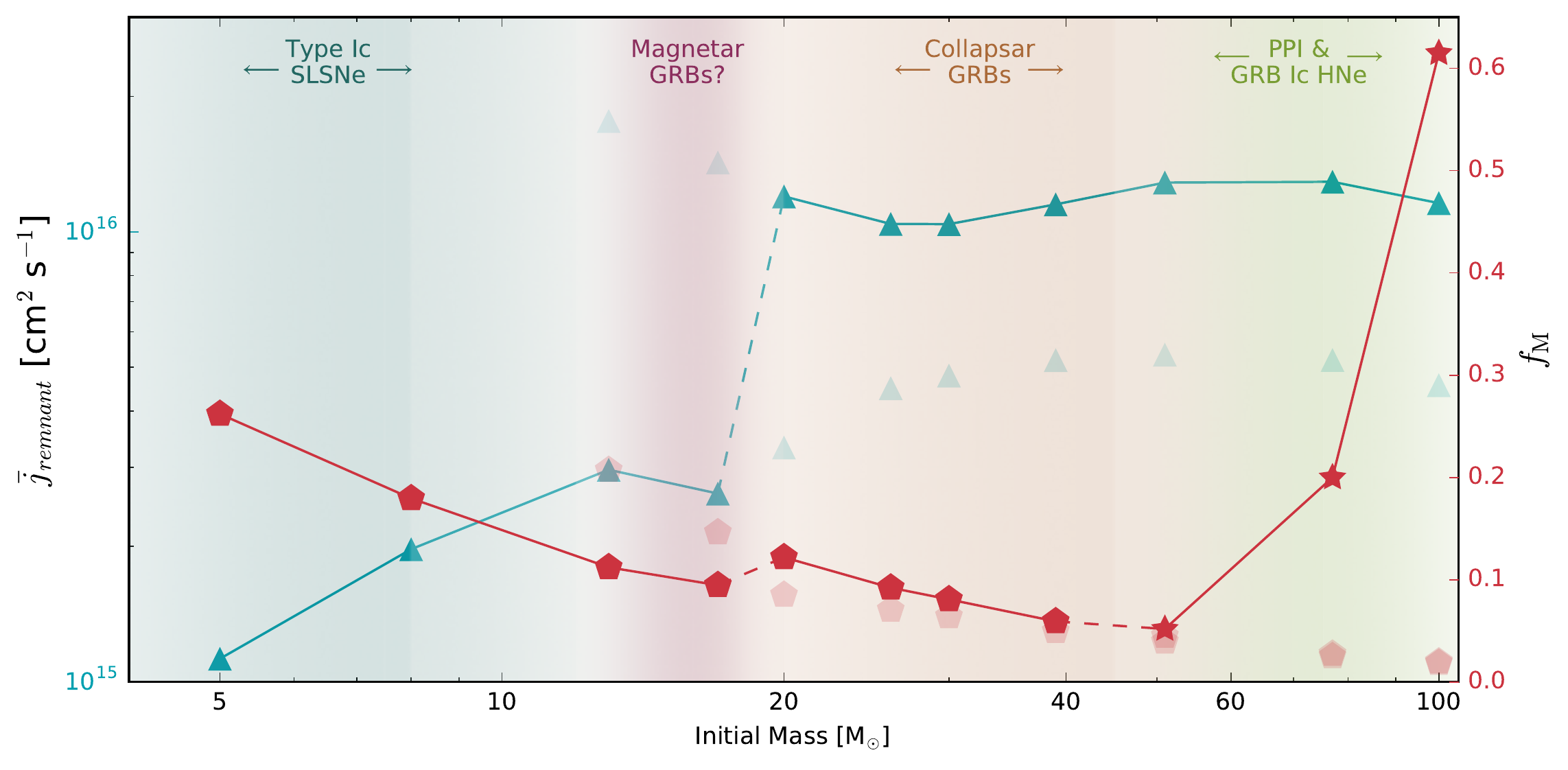}
\caption{Average angular momentum at core O depletion of the first 1.5 $\mso$, for models with mass M$<20 \ \mso$, and first 5 $\mso$ for the rest (left axis, blue triangles); and the ratio between shell and ejecta masses for the same remnant masses (right axis, red pentagons), as a function of initial mass for Series\,B models. The values for converse masses are added in lower saturation for comparison. The ratio between shell and ejecta mass for models that undergo pulsational pair instability is also calculated (red stars), with ejected shell masses according to \cite{2017ApJ...836..244W}.}
\label{fig:massandj}
\end{figure*}

Whereas the shell velocity $\rm{v}_{\rm sh}$ is likely small for our models, 
such that $f_{\rm v}\simeq 1$ (except for the highest masses; see below), 
we need to estimate the SN ejecta mass 
for our models to evaluate the mass ratio $f_{\rm M}$. Corresponding model calculations do no yet go far enough 
in time to give well-founded theoretical ejecta masses for magneto-rotational supernovae \citep{2017MNRAS.469L..43O}.
However, ejecta mass estimates for observed
SN~Ic-bl
are in part quite large \citep[e.g.,][]{2017MNRAS.469.2498M}.
We therefore assume here that except for a neutron star of $1.5\mso$ or, for higher masses, a black hole of
$5\mso$, the remainder of the star will be expelled in the SN. 
Fig.\,\ref{fig:massandj} shows that with this assumption, the mass ratio $f_{\rm M}$ drops from near 
0.3 for our $5\mso$ sequence to 0.06 at $39\mso$, indicating a corresponding
drop in the maximum possible fraction of SN kinetic
energy to be converted into light through SN-CSM interaction. The steep rise of $f_{\rm M}$
for our highest masses in Fig.\,\ref{fig:massandj} assumes massive shells ejected due to the pulsational-pair instability, and will be discussed in Sect.\,\ref{sub:intermediate}.

\subsection{Lowest masses: magnetar powered Type\,Ic SLSNe}

In a recent analysis of 38 H-poor SLSNe in the scope of the magnetar model, \cite{2017ApJ...850...55N} derive the main parameters of such explosions in terms of
initial magnetar spin period ($1.2\dots 4\,$ms) and B-field ($0.2 \times 10^{14}\dots 1.8\times 10^{14}\,$G), 
ejecta mass ($2.2\dots12.9\mso$), and kinetic energy ($1.9\times 10^{51} \dots 9.8\times 10^{51}$\,erg).
Notably, the magnetar spin and ejecta masses are in nice agreement with our lower mass models.
While our estimates for the initial B-fields of the neutron stars are not reliable, we argued in Sect.\,\ref{sec:bfields} that they might well be of magnetar strength. 
Our lower mass models, may thus be suitable to provide 
Type\,I SLSNe within the magnetar model \citep[e.g.,][]{2010ApJ...719L.204W,Kasen2010}.

We can estimate the initial rotational energy of a neutron star forming from our
$5\mso$, $8\mso$, and $13\mso$ Series\,B models as $2.1\times 10^{51}\,$erg, $3.2 \times 10^{51}\,$erg, 
and $9.1 \times 10^{51}\,$erg (cf. Sect.\,\ref{sec:angu}). 
Assuming the average SLSN\,I kinetic energy of $4\times 10^{51}\,$erg (see above), radiation with an energy of up to $\sim 10^{51}\,$erg could in principle be liberated by the interaction of the SN ejecta with
the rotationally driven pre-SN mass loss of about $1\mso$ (Sect.\,\ref{sub:structure}). A larger than canonical 
SN kinetic energy could indeed be plausible since magneto-rotational effects would be expected 
to be very significant here. 

\begin{figure}[ht!]
\gridline{\fig{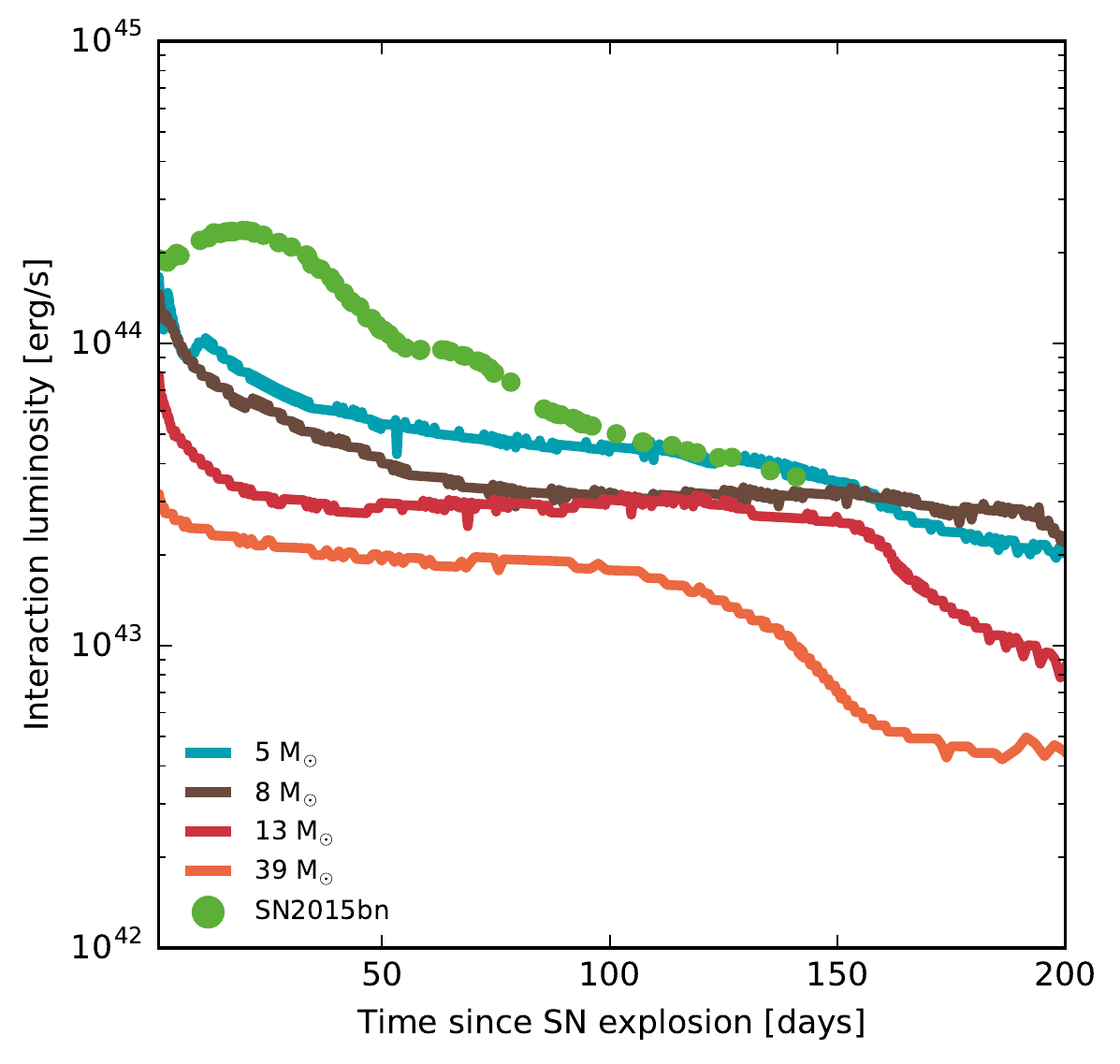}{0.4\textwidth}{}}
\caption{Interaction luminosity adopting a typical SLSN\,I kinetic energy of $4\times 10^{51}\,$erg, calculated using Eq.\ref{eq:deltaE}, and assuming 
the CSM material has spread with a small velocity of $\rm{v}_w = $ 0.2 km/s (see text),
for four Series\,B models with the indicated initial masses. The observed light curve of SN2015bn (from \cite{2016ApJ...828L..18N}) was included for reference, and is displayed with a 10 day shift with respect to its maximum luminosity. }\label{fig:lumis}
\end{figure}

Compared to the rotational energy of a magnetar, the interaction luminosity is smaller but not negligible. 
To demonstrate this, we have computed a hypothetical luminosity that can be released from CSM interaction in our models, assuming
the lost matter has continuously been shed in a spherically symmetric configuration, and with a small constant velocity  ($v_w =$ 0.2\,km\,s$^{-1}$),
with a density profile given by $\rho(r) =
\dot{M}/(4\pi r^2 v_w)$. Adopting the simple inelastic collision model (Eq.\,\ref{eq:fs}), we estimate the liberated interaction luminosity as a function of time for
our lowest mass models by calculating the change in kinetic energy as the ejecta decelerates due to the interaction (Fig.\,\ref{fig:lumis}). We take into account the change in kinetic energy at each time step and assume that all the released kinetic energy is converted to radiation.
This method conserves the momentum of the ejecta and takes in account the change in kinetic energy as it grows in mass. We compare this with the Type\,I SN 2015bn, arguing that the luminosity produced by the interaction should be significant even if the main source of luminosity is the central engine.  

Our estimate of the produced amount of light from the CS interaction does
not assume a complete thermalisation of the SN kinetic energy, but accounts for the
constraints imposed by momentum conservation. Whereas it provides an upper limit,
since even this reduced amount of kinetic energy will not be all 
tansformed into light, \cite{2013MNRAS.428.1020M} showed that it is expected
to match the true efficiency within an order of magitude.
Furthermore, since the CSM mass distribution of our models is not well known, and likely aspherical, Fig.\,\ref{fig:lumis}
should not be taken as a light curve prediction from our models, but only as an
illustration of the potential effects which the interaction may have on the SN light curve.
Hydrodynamical effects and the diffusion of radiation through the dense CSM will shape the actual light curve, but a detailed light curve calculation is beyond the scope of this paper.

It was recently suggested that some external energy source, 
most likely from CSM interaction, may be required to explain various features of SLSNe~Ic.
Many of them show prominent bumps in the declining phase of their light curves
\citep{Inserra2017}, and Fig.\,\ref{fig:lumis} shows that the pre-SN mass loss from our
models may be a candidate to explain those. Furthermore, \citet{Chen2016} find evidence 
for additional energy input in the period of 2\,d before until 22\,d after maximum light
from the spectral evolution of the Type~Ic SLSN\,LSQ14mo, which they tentatively advocate to circumstellar interaction.

The 3D distribution of the mass around the progenitor, with mass likely concentrated towards the equatorial plane, and the temporal behavior of the mass loss in our models may play a role in shaping the SN observables. Short episodes of intense mass loss, such as in our 8 M$_{\odot}$ evolutionary sequence, may lead to dense mass shells, which could give rise to significant ups and downs in the light curve.  \citet{2012ApJ...756L..22M} pointed out that
detached shells necessarily give rise to prominent
``dips'' in the light curve, which is often observed in SLSNe Ic \citep{2016MNRAS.457L..79N}.

We conclude that while the major brightness of magnetar driven Type\,Ic SLSNe is expected to be provided
through heating from the magnetar spin-down \citep{Kasen2010,2010ApJ...719L.204W}, 
significant deviations from the simple magnetar powered light curves  
are expected due to the interaction between ejecta and the CSM, based on our pre-SN models.

\subsection{Intermediate masses: magnetar-driven Type\,Ic GRB SNe?}\label{sub:intermediate}

From the properties of our models, we are not able to  draw the mass limit 
between neutron star and black hole formation. Even for non-rotating massive stars,
this mass limit is not yet well know, since multi-dimensional effects are essential
in core collapse supernovae \citep[e.g.,][]{2016ARNPS..66..341J}.
Parametrised 1D core collapse simulations 
even suggest some stochasticity in the outcome of core collapse within a certain mass range,
rather than a clearcut mass limit (e.g., \citealt{2016ApJ...821...38S}).
Whether the situation is similar in the case of a magneto-rotational core collapse is
unclear. However, black hole formation above a certain mass threshold is expected also in this case
\citep{2017MNRAS.469L..43O}. This provides the major feature dividing explosions which are magnetar powered
from those which are powered by black hole accretion. Therefore, keeping in mind that the situation may 
be more complex, we remain at discussing certain mass ranges for certain outcomes.

In this spirit, it may also be possible that magnetar driven GRBs and associated hypernovae
are produced near the BH/NS mass limit.
In fact, within the scenario of \cite{1993ApJ...408..194T} for B-field production during core collapse, 
the highest B-fields would be expected for the fastest rotating neutron stars, which we find
near the BH/NS mass limit. There, the neutron star masses may also be largest, which may lead to
the most rapidly rotating neutron stars. \citep{2015MNRAS.454.3311M} argue that for
sufficiently strong B-fields, a pulsar rotation period near 2\,ms may lead to a magnetar driven GRB.
Tab.\,\ref{tab:data2}) and Fig. \ref{fig:massandj} show that this may be within reach for our progenitor
models with initial masses below but near $20 \,\mso$. This scenario is attractive because
it can possibly account for ultra-long GRBs as the one reported by \cite{2015Natur.523..189G}. 

\subsection{High masses: collapsar driven Type\,Ic GRB SNe}

As discussed above, we can expect black holes rather than neutron stars being formed above
a certain mass limit, even when rotation and magnetic fields are strongly affecting the iron core collapse
\citep{2017MNRAS.469L..43O}.
Since the black holes are significantly more massive than neutron stars, and as the specific angular momentum
in our models generally increases outwards, their average specific angular momentum will be
larger than that of the neutron stars in the lower mass models. This effect is augmented by a significant drain
of core angular momentum in the lower mass stars due to their relatively large lifetime after core helium 
exhaustion (Sect.\,\ref{sec:angu}). 
As seen in Fig.\,\ref{fig:massandj}, when we assume a black hole mass of $5\mso$, 
their specific angular momenta exceed $10^{16}\,$cm$^2$s$^{-1}$, and the models are thus thought to be suitable 
progenitor models for lGRBs \citep{2005A&A...443..643Y,2006ApJ...637..914W,Yoon2006}. 

lGRBs are associated with hyperenergetic Type\,Ic SNe. 
Our more massive models have larger ejecta masses, and are expected 
to convert only up to about 5\% of the SN kinetic energy into radiation through CSM interaction. 
These supernovae have kinetic energies
of the order of $10^{52}\,$erg, but they are not superluminous, and only few of them have been observed
in significant detail until today (see, for example, \citealt{2017AdAst2017E...5C}). 

At face value, our estimate would imply that up to $5\,\times 10^{50}\,$erg of radiation 
could be produced through CSM interaction in GRB SNe. 
However, we have strong evidence that these hypernova explosions are strongly aspherical \citep[e.g.,][]{2002ApJ...565..405M}, with much higher
ejecta velocities in polar than in equatorial direction. At the same time, it is likely that the
pre-supernova mass loss of our model stars would be emitted predominantly into directions of low latitude.  
The hypernova might thus catapult its fastest ejecta into the polar holes of the CSM distribution.
The true interaction luminosity could therefore be far below the number quoted above.
Still, around $1\,\mso$ of material is expected to sit close to the exploding star and,
depending on the viewing angle, it may show up in absorption or emission.

Also the GRB jet is expected to be launched in the polar direction. As the densest CSM is likely located in the equatorial plane, an interaction between the jet and the CSM may not occur often. However, several low-luminosity lGRBs like GRB060218/SN~2006aj \citep[e.g.,][]{Campana2006,Mazzali2006,Pian2006,Soderberg2008} are suggested to
occur in a dense CSM surrounding the progenitor stars \citep[e.g.,][]{Irwin2016}.

\subsection{Very high masses: interaction powered Type\,Ic superluminous supernovae}

At the highest masses considered here, above 50 $\mso$, the centrifugally suported ejection of about 
$1\,\mso$ of material is perhaps unimportant. However, the oxygen cores of the considered models become 
e$^{\pm}$-pair-unstable
before oxygen ignition, and may undergo pair instability pulses
\citep{2007Natur.450..390W,2012ApJ...760..154C,2017ApJ...836..244W,2016MNRAS.457..351Y}. In non-rotating models,
\cite{2017ApJ...836..244W} has shown these pulses may eject shells with masses up to 28$\,\mso$, with higher mass
shells for more massive stars. Notably, the models which produce the so called pulsational pair-instability SNe
will eventually undergo iron core collapse. Assuming that during the pair-induced pulsations the core
angular momentum remains unchanged, this core collapse may produce a lGRB, or at least an engine-driven
hypernova. It may therefore be a common outcome from our most massive models to produce 
an interaction powered SLSN of Type\,Ic, where the interaction occurs between a massive 
Type\,Ic hypernova and a massive circumstellar shell consisting predominantly of carbon and oxygen. However, we note that 
the effect of the pulsations on the core
angular momentum distribution remains yet to be determined by detailed models.

Within this scenario, depending on the uncertain mass of the black hole produced in the collapse,
the conversion efficiency of supernova kinetic energy to light may be very high (Fig. \ref{fig:massandj}). 
With the central engine
producing a hypernova, the kinetic energy of the ejecta would also be very high. 
As the ejecting force which lifts off the massive shell during a pair-induced pulse is
rooted deep inside the star, where ratio of centrifugal force to gravity is well below one,
the circumstellar shell produced in this event is expected to be far less aspherical than
the wind-produced shells discussed above. Therefore, even though the high velocity
of the shell produced by the pair pulsation may lead to $f_{\rm v}$-values below one
(cf., Eq.\,\ref{eq:fs}), this
scenario may potentially produce the brightest interaction driven supernovae, with a total energy content of
the radiation of the order of $10^{52}\,$erg, competing with the brightest possible 
magnetar driven supernova \citep{2016ApJ...820L..38S}.

PPI supernovae may in principle also produce interaction dominated Type\,Ic supernovae
if they would not rotate. However, according to the current core collapse supernova models
(\citealt{HegerWoosley2002}, \citealt{2016MNRAS.456.1320T}), the final collapse of the core into a black hole in non-rotating stars of such high mass may actually not produce a core collapse supernova at all.

It may be interesting to compare our model predictions with the SLSN\,Ic Gaia2016apd.
This supernova is peculiar as it shows an extremely blue continuum, much more so
than any other of the generally already very blue Ic\,SLSNe \citep{2017ApJ...840...57Y}.  \cite{2017ApJ...835L...8N} explain this
as a magnetar powered supernova with rather ordinary magnetar parameters 
($P_{\rm }=1.9\,$ms, and $B=2\,10^{14}\,$G) but with a very strong central engine,
giving rise to a SN kinetic energy of at least $3\times 10^{51}\,$erg, possibly
$> 10^{52}\,$erg. \cite{2017ApJ...845L...2T}, however, by employing multi-group
radiative transfer calculations, do not succeed to explain the extremely blue continuum with
this scenario. They find a hypernova interacting with a fast and massive
CO-shell as most likely explanation for SN Gaia2016apd. We note that this at first glance very exotic scenario is a natural prediction from our most massive progenitor models. 

At even higher masses than considered in our work, we would expect plain pair instability
supernovae \citep{HegerWoosley2002,2014A&A...565A..70K,2017MNRAS.464.2854K,2011ApJ...734..102K,2013MNRAS.428.3227D}.
Whereas the influence of rotation on PISNe has been investigated \citep{1985A&A...149..413G,ChatzopolousWheeler2012, 2012ApJ...760..154C,2013ApJ...776..129C},
the magneto-rotational PI-collapse remains yet to be studied.

\section{Discussion} \label{sec:discussion}

The results derived above have been obtained through stellar evolution calculations in the frame
of the CHE scenario \citep{2005A&A...443..643Y,2006ApJ...637..914W}, in which we artificially
increased the diffusion coefficient for rotationally induced mixing. For this diffusion coefficient,
and its dependence on the stellar rotation rate, only local and linear approximations are known
(cf., \citealt{HegerLangerWoosley2000,2002A&A...381..923S}), while the non-linear behavior, which might 
be particularly relevant in the case of extreme rotation considered here, is not known 
from first principles.

While used ad-hoc, some retrospect support for the enhancement of the mixing coefficient may be drawn from the fact that
the corresponding models do reproduce Type\,Ic supernova progenitors, which has proven notoriously 
difficult for CHE models using the nominal diffusion coefficient, in particular in the lower mass range
of the models considered here \citep{Yoon2006}. We emphasise that this does not alleviate the need to find confirmation
for the faster mixing, through multidimensional MHD calculations as well a through observations.

In any case, our calculations produce rapidly rotating nearly bare CO cores at the end of core He
burning, and the observations of GRB supernovae and Type\,Ic SLSNe tell us that such objects may indeed be
required to explain what we see. While binary evolutionary channels have been proposed to produce
such objects (e.g., \citealt{2008A&A...484..831D,2010MNRAS.406..840P,2014ApJ...793L..36F,Marchant2016}), we
point out that once they are formed, their further evolution will be rather independent of the the formation channel.
The reason is that the chemical structure is uniquely defined --- they are just CO stars ---,
and at least in the framework of the Spruit-Taylor dynamo the angular momentum distribution will 
also be homogeneous, as at this stage there are no entropy barriers inside the stars. 

We therefore suggest that whether CHE is the dominant production channel or not,
and independent of whether the mixing coefficient is really as large as we assumed, 
our evolutionary models do describe the post-core He burning evolution of the progenitors 
of GRB supernovae and Type\,Ic SLSNe. I.e., the overall contraction and spin-up induced mass loss,
as well as the mass dependence of the specific angular momentum of the compact remnant
(Fig.\,\ref{fig:massandj}), are not expected to depend significantly
on the formation history of the rapidly rotating CO stars.

When considering the spin-up induced mass loss a generic prediction of the magnetar and collapsar 
models discussed here, the diagnostics of vigorous pre-SN mass loss and circumstellar shells 
in the observations of related supernovae becomes crucial, in the sense that
a failure to find evidence of such mass loss may question the underlying model.
However, as discussed in Sect.\,\ref{sec:consequences}, even though such diagnostics are often difficult to obtain, 
many indications for significant CSM interaction have already been found, at least in SLSNe.

At the same time, the detection of large amounts of CSM close to the exploding star in a Type\,Ic
supernova would not prove the spin-up scenario. Other ways of achieving this have been proposed,
e.g., through non-conservative close binary mass transfer after core-He exhaustion \citep{2012ApJ...752L...2C}, 
through the transport of energy by gravity waves excited in the deep interior during the
late burning stars from the core to the stellar surface \citep{2017arXiv171004251F}, or through
violent pulsations \citep{1997A&A...327..224H,2010ApJ...717L..62Y,2015A&A...573A..18M}. In contrast to these mechanisms,
the one discussed here works only in extremely rapidly rotating stars and thus in rotation
powered supernovae. In that sense, the indications for CSM interaction in a large number of
Type\,I SLSNe is supporting the scenario of a rapidly rotating progenitor, independent 
of the need for this within the magnetar scenario.

In this work, we have computed only models with an extremely small metallicity ($Z_{\odot}/50$). While 
massive stars with such low metallicities exist in the local universe \citep{2016MNRAS.457...64I}, and
Type\,I SLSNe \citep{2017MNRAS.470.3566C,2018MNRAS.473.1258S} and lGRBs 
\citep{2017ApJ...834..170G} do preferentially occur at sub-solar metallicity,
the observationally deduced metallicities are generally larger than $Z_{\odot}/50$.

We have chosen such a low metallicity, since from the calculations of \cite{Yoon2006} and \cite{Marchant2016}
which explored CHE at various metallicities, it was clear that this would open the maximum mass range for this
scenario, and --- as discussed above --- once a rapidly rotating CO star is produced, its progenitor evolution
does not affect its final evolution. Indeed, as ordinary stellar winds are too weak to affect the post-core
helium burning evolution, the evolution through the late burning stages will also be largely metallicity-independent. 

However, within the CHE production channel of Type\,I SLSNe and lGRBs, metallicity will foreseeably strongly affect
the formation rates of these events, mediated through the metallicity dependance of radiation driven winds
\citep{2007A&A...473..603M,2006astro.ph..6580C}.
In fact, to the extent that even in
metal-rich stellar surfaces, iron is the main element driving the wind \citep{2005A&A...442..587V}, the
metallicity dependance of the final core angular momentum in CHE models computed with enhanced mixing coefficients
is expected to be very similar for the highest masses to that obtained by 
\cite{Yoon2006}.
This is confirmed by comparing the numbers in  Tabs.\,\ref{tab:data2} and\,\ref{tab:data} with those obtained by \cite{Yoon2006}
for their lowest considered metallicity (Z=0.00001). Therefore, the main effect of metallicity in our models
is expected to be that higher metallicity lowers the mass threshold above which the CHE scenario 
can not work any more due to wind induced spin-down during H- and He-burning (see also \citealt{Marchant2016}).
Grids of detailed models are required to confirm this, and to compare with the empirical magnetar spin-metallicity
correlation suggested by \cite{2017MNRAS.470.3566C} although the correlation might not be as strong as originally suggested \citep[e.g.,][]{2017ApJ...850...55N,2017arXiv170801623D}.

One might expect from our models that the low metallicity bias of SLSNe\,I 
may be smaller than that of lGRBs, since originating from a lower mass range implies
that wind spin down is less important. Currently, observations seem to imply the opposite
trend \citep{2017ApJ...834..170G,2017MNRAS.470.3566C}. However, \cite{Yoon2006}
found that the lower mass limit for chemically homogeneous evolution increases with 
metallicity, which could strengthen the low metallicity bias for SLSNe\,I.
Model grids with varying metallicity are urgently needed to enlighten this point.

In the same spirit, one might consider the frequency of the the various types
of explosions expected from our models in connection with the predicted 
corresponding mass ranges, assuming an initial mass function. However, as the numbers
might change considerably when also a convolution with the respective metallicty
trends would be included, we refrain from such an attempt here, but point out that
future model grids with varying metallicity would allow to do this.
The only statement we may make in this direction is that the outcome expected for the most massive stars considered here, the superluminous interacting Ic hypernova, is 
perhaps indeed the observationally rarest case (as pointed out by our referee).

\section{Conclusions}\label{sec:conclusions}

We obtain fast spinning, almost pure CO\,stars at core helium exhaustion, with high angular momenta, corresponding to progenitors of SLSNe, Type Ic SNe and GRBs, computed through evolutionary sequences for rapidly rotating low metallicity massive stars that evolve quasi-chemically 
homogeneously. Our models using the nominal rotational mixing efficiency correspond well to earlier such calculations
\citep{2005A&A...443..643Y,2006ApJ...637..914W,Yoon2006,Szecsi2015}. When
invoking enhanced rotational mixing, we obtain rapidly rotating CO\,stars.
We argue that these represent the progenitors of Type\,Ic SLSNe and lGRBs, independent of the evolutionary
scenario through which these explosions are actually produced, in as much as rapid rotation and the absence of
helium are observational requirements.

We follow the evolution of these objects through the late burning stages until the pre-collapse state. We find that 
due to the lack of a helium envelope, they contract and do so at an accelerated rate, following the neutrino-mediated
Kelvin-Helmholtz timescale (Eq.\,\ref{eq:kh}). The corresponding spin-up leads to centrifugal shedding of significant amounts
of mass during the last hundreds of years of the stars evolution, which will affect the display of the ensuing 
supernova explosion. We further find a trend of increasing specific angular momentum in the innermost cores 
with mass. Together with the fact that our mass range covers the NS/BH as well as the BH/pair-instability transition,
this leads to the suggestion of the following explosion types as function of mass. 

For the lowest considered masses, with ejecta masses in the range $2\mso\dots 20\mso$, our models predict 
NS spin rates as required for magnetar-powered Type\,Ic SLSNe \citep{2015MNRAS.454.3311M}, and with magnetic fields that are consistent with those derived to reproduce the light curves of SLSNe \cite{2017ApJ...850...55N}. In addition to the magnetar-heating
in such events, our models show that in addition, up 20\% of the SN kinetic energy could be converted into
radiation through CSM interaction. Near the BH formation mass limit, magnetar-driven (ultra-)long GRBS may 
also be possible.

Above the BH formation threshold, the average specific angular momentum of the compact object reaches 
$\sim 10^{16}\,$cm$^2$s$^{-1}$, and lGRBs are expected within the collapsar scenario \citep{Woosley1993}, accompanied
by Type\,Ic hypernovae (i.e., supernovae with a large kinetic energy). The predicted mass of the close range CSM amounts here to a few percent 
of the mass of the hypernova ejecta, with an efficient thermalisation potentially prevented by geometry effects.

The oxygen cores of the largest mass stars considered here become e$^{\pm}$-pair unstable (the last stage considered
by our models), and are expected to eject massive shells due the pulsational pair instability \citep{2017ApJ...836..244W}. 
As for the lower mass cases, the central engine is expected to produce a hypernovae, which here 
would collide with the massive CO-shells ejected earlier to produce an interaction dominated
Type\,Ic SLSN with up to $10^{52}\,$erg of energy in the form of light. 

The properties of the explosion phenomena discussed above are anticipated based upon our current knowledge.
Therefore, besides exploring the pre-supernova model parameter space in greater depth, multi-D models of
of the pre-SN CSM evolution, and of the magneto-rotational collapse phase would be valuable next steps.
This would allow for consistent calculations of the resulting light curves and spectra of the supernova
phase, which are ultimately needed as a test of the models presented here.  

\acknowledgments
Acknowledgments. 
We would like to acknowledge valuable discussions with Sergei Blinnikov, Pablo Marchant, Thomas Tauris and Alexey Tolstov, and very usefull comments from the anonymous referee of our paper. We would like to thank Matt Nicholl for providing the data for SN2015bn.

\software{MESA (v10000; \cite{MESAI,MESAII,MESAIII,2017arXiv171008424P})}

\bibliographystyle{yahapj}
\bibliography{references}

\appendix
\section{Additional information}

\begin{figure}[ht!]
\gridline{\fig{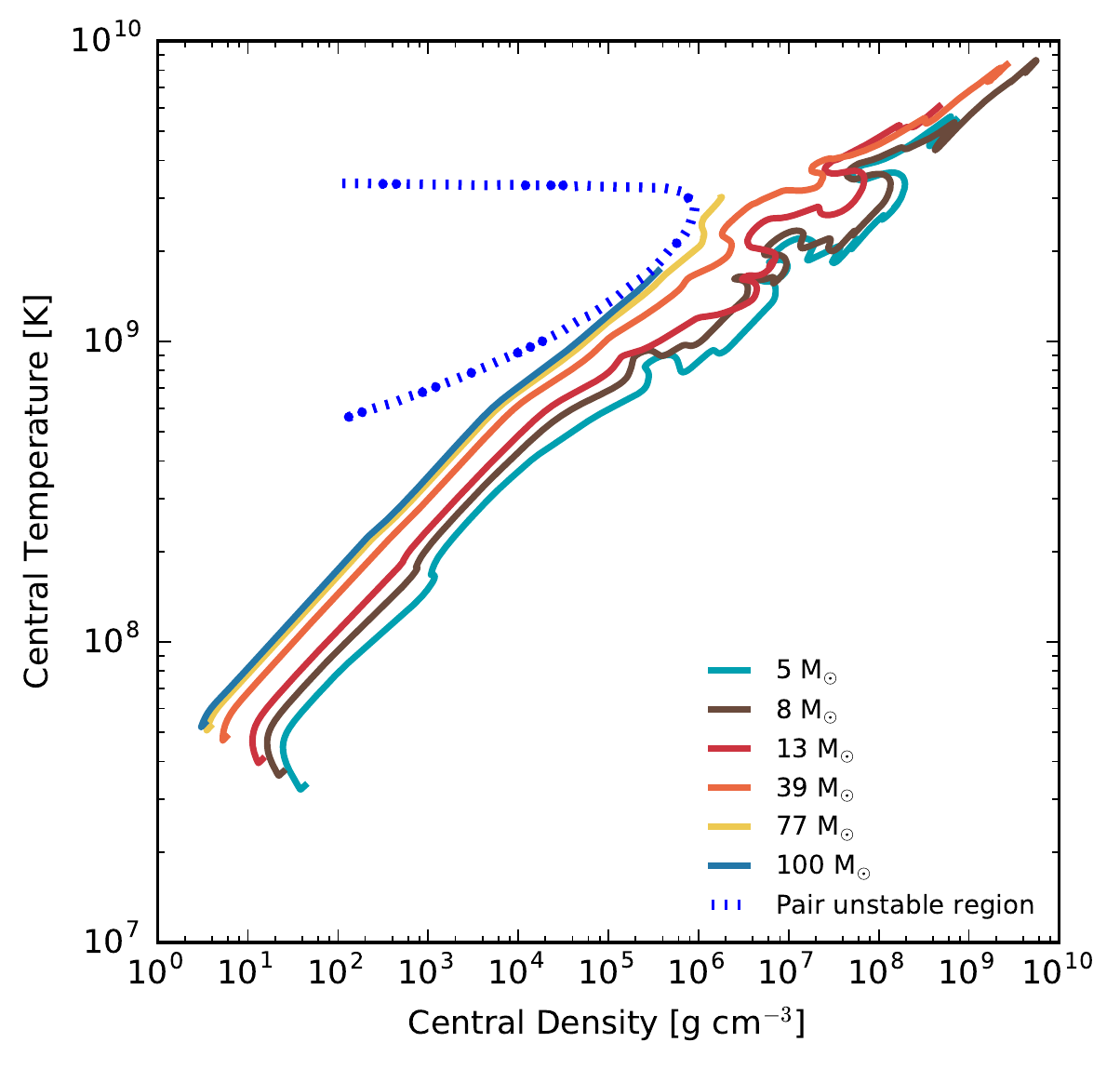}{0.4\textwidth}{}}
\caption{Central temperature versus central density evolution for Series\,B models of the indicated initial masses.}\label{fig:rhoc-Tc}
\end{figure}

\begin{figure}[ht!]
\gridline{\fig{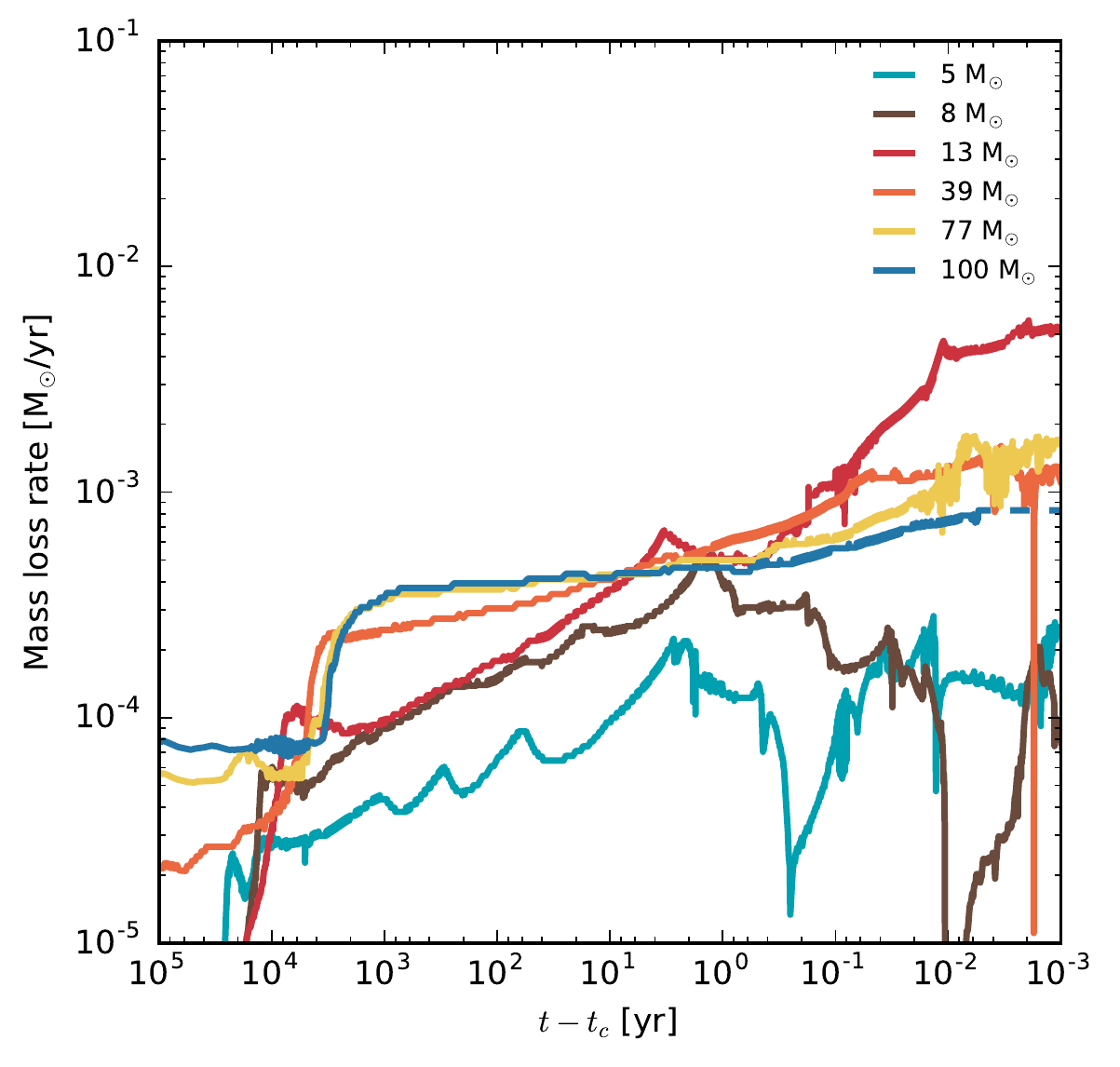}{0.4\textwidth}{}}
\caption{Mass loss rates as a function of remaining time for Series\,B models}\label{fig:mdotvst}
\end{figure}

In this section we provide additional information about our models. For Series\,B models, we include Fig. \ref{fig:rhoc-Tc}, which presents the evolution of the central densities and temperatures, Fig. \ref{fig:mdotvst}, which complements Fig.\,\ref{fig:mdot} in showing the mass loss rate as a function of remaining time.

\begin{figure*}[ht!]
\plotone{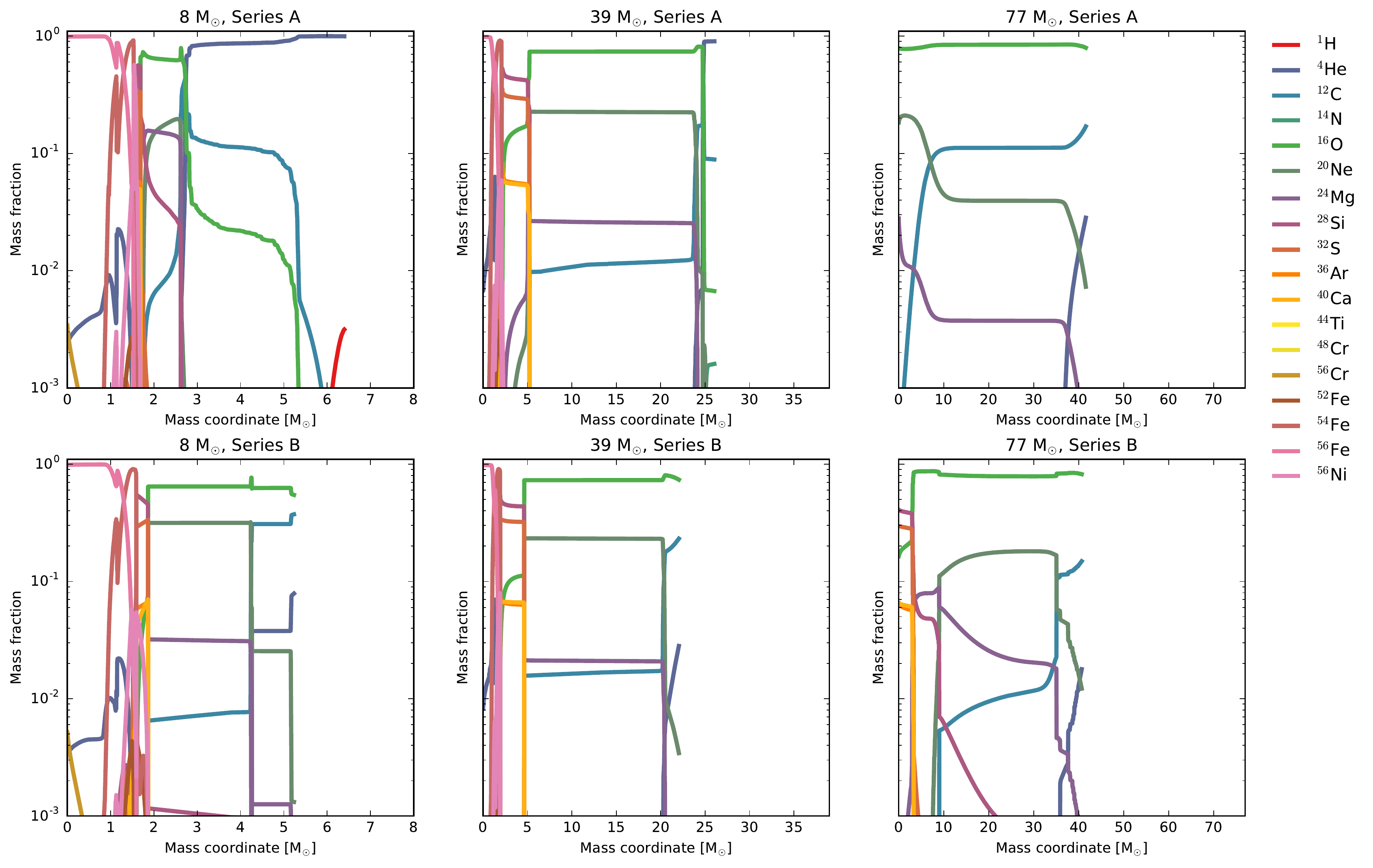}
\caption{Mass fractions of the chemical elements as function of
the Langrangian mass coordinate in six different representative pre-SN stellar models.}\label{fig:abundancesfin}
\end{figure*}

Fig.\,\ref{fig:abundancesfin} presents the abundances of different elements for selected pre-SN models both from Series\,A and Series\,B.

For Series\,A, we include Tab.\,\ref{tab:data}, with relevant values, complementing Tab.\,\ref{tab:data2}.

\begin{deluxetable*}{c|cccccc}[ht!]
\tablecaption{Initial and final parameters of the evolutionary sequences in Series\,A.\label{tab:data}}
\tablecolumns{7}
\tablewidth{0pt}
\tablehead{
\colhead{Initial mass [M$_{\odot}$]} & \colhead{5}  & \colhead{8}  & \colhead{13}  & \colhead{39}  & \colhead{77}  & \colhead{100} 
}
\setlength\extrarowheight{-1.6pt}
\startdata
Initial $\Omega / \Omega_{\rm{crit}}$ & 0.89 &  0.84 &  0.79 &  0.69 &  0.67 &  0.68 \\
CO core mass [M$_{\odot}$]  \tablenotemark{a} & 1.43 & 2.63 & 6.06 & 24.80 & 42.66 & 53.51 \\
Final mass [M$_{\odot}$] & 4.19 & 6.40 & 9.61 & 26.04 & 41.59 & 52.47 \\
$\Delta \rm{M}_{\rm H \rightarrow He}$ [M$_{\odot}$] \tablenotemark{b} & 0.51 & 0.58 & 1.05 & 4.48 & 13.21 & 19.84 \\
$\Delta \rm{M}_{\rm He \rightarrow final}   $[M$_{\odot}$]   \tablenotemark{c}   & $ 0.01 $  & $ 0.25 $& $ 0.83 $  & $ 1.27 $  & $ 1.06 $  & $ 1.00 $ \\
$f_{\rm M}$ & $ 0.004 $  & $ 0.05 $& $ 0.09 $  & $ 0.05 $  & $ 0.20 $  & $ 0.60 $ \\
Final Radius [cm] & $ 2.38 \times 10^{11} $  & $ 1.76 \times 10^{11} $& $ 1.34 \times 10^{11} $  & $ 3.43 \times 10^{10} $  & $ 3.90 \times 10^{10} $  & $ 4.62 \times 10^{10} $ \\
Final T$_{\rm{eff}}$ [K] & $ 4.85 \times 10^{4} $  & $ 6.86 \times 10^{4} $& $ 9.52 \times 10^{4} $  & $ 2.12 \times 10^{5} $  & $ 2.86 \times 10^{5} $  & $ 2.84 \times 10^{5} $ \\
Final H mass [M$_{\odot}$] & $ 0.01 $  & $ 0$& $ 0 $  & $ 0 $  & $0 $  & $ 0 $ \\
Final He mass [M$_{\odot}$] & $ 2.32$ & $ 3.33$ & $ 3.19  $ & $ 1.16 $ & $ 0.05$ & $ 0.07 $ \\
Final Y$_{\rm{surf}}$ & $ 0.98  $ & $ 1.00 $ & $ 0.99 $ & $ 0.90  $ & $ 0.03$ & $ 0.03$ \\
Radius at $\tau$ = 1 [cm] & $ - $  & $ - $& $ - $  & $ 9.05 \times 10^{13} $  & $ 1.61 \times 10^{13} $  & $ 1.22 \times 10^{13} $ \\
$\bar{j}_{\rm 1.5 \mso}$ [cm$^2$ s$^{-1}$] &  $ 2.46 \times 10^{14} $   & $ 2.83 \times 10^{14} $& $ 8.88 \times 10^{14} $  & $ 1.88 \times 10^{15} $  & $ 5.20 \times 10^{15} $  & $ 4.48 \times 10^{15} $ \\
$\bar{j}_{\rm 2 \mso}$ [cm$^2$ s$^{-1}$]  & $ 8.32 \times 10^{14} $ & $ 3.57 \times 10^{14} $ & $ 8.88 \times 10^{14} $ & $ 3.72 \times 10^{15} $ & $  6.41 \times 10^{15} $ & $  5.73 \times 10^{15}$ \\
$\bar{j}_{\rm 5 \mso}$ [cm$^2$ s$^{-1}$]   & -- & $ 2.73 \times 10^{16} $ & $ 3.40 \times 10^{15} $ & $ 1.12 \times 10^{16} $ & $ 1.31 \times 10^{16} $ & $1.15 \times 10^{16}$ \\
$P_{\rm rot}$ (1.5 $\mso$ NS) [ms] \tablenotemark{d} & 19.42 & 16.89 &  5.39 &  2.54 &  0.92 &  1.07 \\
$E_{\rm rot}$ (1.5 $\mso$ NS) [erg] \tablenotemark{d}&$ 1.19 \times 10^{50} $ & $ 1.57 \times 10^{50} $ &  $ 1.54 \times 10^{51} $ &  $ 6.93 \times 10^{51} $ &  $ 5.29 \times 10^{52} $ &  $ 3.92 \times 10^{52} $\\
Final fate &  CCSN  & CCSN &  CCSN  & CCSN   &  PPISN   &  PPISN  \\ 
\enddata
\tablenotetext{a-d}{\, \, \, \, See Tab.\,\ref{tab:data2}}
\end{deluxetable*}

\end{document}